\input harvmac

\def\Title#1#2{\rightline{#1}\ifx\answ\bigans\nopagenumbers\pageno0
\vskip0.5in
\else\pageno1\vskip.5in\fi \centerline{\titlefont #2}\vskip .3in}

\font\caps=cmcsc10

\noblackbox
\parskip=1.5mm

  
\def\npb#1#2#3{{\it Nucl. Phys.} {\bf B#1} (#2) #3 }
\def\plb#1#2#3{{\it Phys. Lett.} {\bf B#1} (#2) #3 }
\def\prd#1#2#3{{\it Phys. Rev. } {\bf D#1} (#2) #3 }
\def\prl#1#2#3{{\it Phys. Rev. Lett.} {\bf #1} (#2) #3 }
\def\mpla#1#2#3{{\it Mod. Phys. Lett.} {\bf A#1} (#2) #3 }
\def\ijmpa#1#2#3{{\it Int. J. Mod. Phys.} {\bf A#1} (#2) #3 }

\def\cmp#1#2#3{{\it Commun. Math. Phys.} {\bf #1} (#2) #3 }

\def\bb#1{{\tt hep-th/#1}}

\def\rmp#1#2#3{{\it Rev. Mod. Phys.} {\bf #1} (#2) #3 }
\def\jetp#1#2#3{{\it Sov. Phys. JEPT} {\bf #1} (#2) #3 }
\def\pha#1#2#3{{\it Physica } {\bf A#1} (#2) #3 }


           \def\CO{{\cal O}} \def\CZ{{\cal Z}}
   
\def\CL{{\cal L}} \def\CH{{\cal H}}  
  \def\CD{{\cal D}}

\def\CK{{\cal K}}


\def\dj{\hbox{d\kern-0.347em \vrule width 0.3em height 1.252ex depth
-1.21ex \kern 0.051em}}

\def\half{{1\over 2}\,}
\def\d{{\rm d}}

\def\ket{\rangle}
\def\bra{\langle}

\def\balpha{\overline \alpha}
\def\bS{\overline S}

\def\pt{\partial}

\lref\rgreenf{M.B. Green, \plb {329}{1994}{435} (\bb{9403040}).}
\lref\rpolsem{J. Dai, R.G. Leigh and J. Polchinski, \mpla{4}
 {1989}{2073.}}
\lref\rpolrr{J. Polchinski, \prl{75}{1995}{4728} (\bb{9510017}).}
\lref\rgreent{M.B. Green, \plb {282}{1992}{380} (\bb{9201054}).}
\lref\rgreenb{M.B. Green, \plb {266}{1991}{325.}} 
\lref\ranalogo{J. Polchinski, \cmp{104}{1986}{37.}}
\lref\rpoint{E.F. Corrigan and D.B. Fairlie, \npb{91}{1975}{527\semi} 
M.B. Green, \plb{69}{1976}{89\semi}
I.R. Klebanov and L. Thorlacius, \plb {371}{1996}{51}  (\bb{9510200)}\semi
J.L.F. Barb\'on, \plb {382}{1996}{60} (\bb{9601098}).}  
\lref\rcdg{C.G. Callan, R. Dashen and D.J. Gross, \prd {17}{1978}{2717.}}
\lref\raffleck{I. Affleck, \npb {191}{1981}{429.}}
\lref\rbsus{T. Banks and L. Susskind, {\it Brane-antibrane forces}, RU-95-87
 (\bb{9511194}).}
\lref\rpolcom{J. Polchinski, \prd {50}{1994}{6041} (\bb{9407031}).}
\lref\rmolive{C. Montonen and D. Olive, \plb {72}{1977}{117.}}
\lref\rgpy{D.J. Gross, R. Pisarski and L. Yaffe, {\it Rev. Mod. Phys.} 
{\bf 53} (1981) 43.}     
\lref\rahw{I. Affleck, J. Harvey and E. Witten, \npb {206}{1982}{413.}}
\lref\rtd{P. Horava, \plb {231}{1989}{251\semi}
J. Polchinski, S. Chaudhuri and C.V. Johnson, NSF-ITP-96-003, 
\bb{9602052\semi}
E. \'Alvarez, J.L.F. Barb\'on and J. Borlaf, PUPT-96-1601, (\bb{9603089}).}
\lref\rgreengut{M.B. Green and M. Gutperle, DAMPT-96-34, (\bb{9604091}).}
\lref\rbbs{K. Becker, M. Becker and A. Strominger, \npb {456}{1995}{130},
 (\bb{9507158}).}
\lref\ritep{V.A. Novikov, M.A. Shifman, A.I. Vainshtein, V.B. Voloshin 
and V.I. Zakharov, \npb {229}{1983}{394.}}
\lref\rpoly{A.M. Polyakov, \npb {120}{1977}{429.}}
\lref\rsch{J.H. Schwarz, \npb {226}{1983}{269.}}
\lref\rgs{M.B. Green and J.H. Schwarz, \plb {122}{1983}{143.}}
\lref\rggp{G.W. Gibbons, M.B. Green and M.J. Perry, \plb {370}{1996}{37}
(\bb{9511080}).}
\lref\rwit{E. Witten, IASSNS-HEP-96-29, \bb{9604030.}}
\lref\rgreengas{M.B. Green, \plb {354}{1995}{271} (\bb{9504108}).}
\lref\cabpar{N. Cabibbo and G. Parisi, \plb{59}{1975}{67.}} 
\lref\ric{N. Deo, S. Jain and C.-I. Tan, \plb{220}{1989}{125.}}
\lref\rminahan{J. Minahan, \npb{333}{1990}{525.}}
\lref\rcar{R.D. Carlitz, \prd{5}{1972}{3231.}}
\lref\rfra{S. Frautschi, \prd{3}{1971}{2821.}}
\lref\dokph{E. D'Hoker and D.H Phong, \rmp{60}{1988}{917.}}
\lref\rshenker{S. Shenker, {\it The strength of non-perturbative
effects in String Theory}, in: ``Carg\`ese 1990, Proceedings,
Random surfaces and quantum gravity''.}
\lref\raorw{E. \'Alvarez and M.A.R. Osorio, \pha{158}{1989}{449.}}
\lref\rmipaper{J.L.F. Barb\'on, {\it Fermion exchange between D-instantons},
CERN-TH/96-360 ({\tt hep-th/9701075}).}
\lref\rpolrev{J. Polchinski, {\it TASI lectures on D-branes,} NSF-ITP-96-145 
({\tt hep-th/9611050}).}
\lref\rtofito{G. 't Hooft, \prd{14}{1976}{3432.}}
\lref\raloros{E. \'Alvarez, T. Ort\'{\i}n and M.A.R. Osorio, 
\prd{43}{1990}{3990.}}
\lref\rgbu{M.B. Green, \npb{381}{1992}{201.}}
\lref\rform{I.R. Klebanov and L. Thorlacius, \plb{371}{1996}{51}
(\bb{9510200})\semi
J.L.F. Barb\'on, \plb{382}{1996}{60} (\bb{9601098}).}
\lref\rprev{J. Polchinski, \rmp{68}{1996}{1245} (\bb{9607050}).}
\lref\rmalda{J.M. Maldacena, {\it Black holes in string theory}, Princeton
University Ph.D. thesis
(\bb{9607235}).}
\lref\rpolins{J. Polchinski, \prd{50}{1994}{6041} (\bb{9407031}).}
\lref\rmio{J.L.F. Barb\'on, \npb{452}{1996}{313} (\bb{9506137}).}
\lref\rwittb{E. Witten, \npb{460}{1996}{335} (\bb{9510135}).}
\lref\rgg{M.B. Green and M. Gutperle, {\it Effects of 
D-instantons}, DAMTP-96-104  (\bb{9701093}).}
\lref\rseishe{N. Seiberg and S. Shenker, \prd{45}{1992}{4581} (\bb{9201017}).}
\lref\renbor{E. \'Alvarez, J.L.F. Barb\'on and J. Borlaf, \npb{479}{1996}{218}
(\bb{9603089}).}

\lref\dob{C.P. Burgess and T.R. Morris, \npb{291}{1987}{256;} 
\npb{291}{1987}{285\semi}
A. Morozov and A. Rosly, \plb{214}{1988}{522;} \npb{326}{1989}{185;}
\npb{326}{1989}{205\semi}
I.D. Vaisburg, \mpla{3}{1988}{511.}}
\lref\blau{S.K. Blau, M. Clements, S. Della Pietra, S. Carlip and V. Della
Pietra, \npb{301}{1988}{285.}}
\lref\rdual{M.A. V\'azquez-Mozo, \plb{388}{1996}{494} (\bb{9607052}).}
\lref\alos{E. \'Alvarez and M.A.R. Osorio, \prd{36}{1987}{1175.}}
\lref\ralor{E. \'Alvarez and T. Ort\'{\i}n, \plb{241}{1990}{215.}}
\lref\rena{E. \'Alvarez, \npb{269}{1986}{596.}}
\lref\moore{G. Moore, \plb{176}{1986}{369.}}
\lref\rbn{A.A. Belavin and V.G. Knizhnik, \jetp{64}{1986}{214.}}
\lref\rostr{A. Sagnotti, \prep{184}{1989}{167\semi}
P. Horava, \plb{231}{1989}{251;} \npb{327}{1989}{461\semi}
M. Bianchi and A. Sagnotti, \plb{247}{1990}{517\semi}
M. Bianchi, G. Pradisi and A. Sagnotti, \npb{376}{1992}{365.}
}
\lref\rgreen{M.B. Green, \npb{381}{1992}{201.}}
\lref\ratickw{J. Atick and E. Witten, \npb{310}{1988}{291.}}
\lref\ralos{E. \'Alvarez and M.A.R. Osorio, \npb{304}{1988}{327.}}
\lref\rbranva{R.H. Brandenberger and C. Vafa, \npb{316}{1988}{391.}}
\lref\rpol{J. Polchinski, \cmp {104}{1986}{37.}} 
\lref\rgr{M.B. Green, \plb{282}{1992}{380.}}
\lref\rb{J.L.F. Barb\'on, Preprint PUPT-96-1635, to appear.}
\lref\rrev{J. Polchinski, S. Chaudhuri and C. Johnson, Preprint NSF-ITP-96-003,
hept-th/9602052.}
\lref\rao{E. \'Alvarez and M.A.R. Osorio, \prd{36}{1987}{1175.}}
\lref\rpcmp{J. Polchinski, \cmp{104}{1986}{37.}}
\lref\rtdual{
K.H. O'Brien and C.I. Tan, \prd {36}{1987}{1184\semi}
B. McClain and B.D.B Roth, \cmp{111}{1987}{539\semi}
E. \'Alvarez and M.A.R. Osorio, \prd{40}{1989}{1150.} }
\lref\rrohm{R. Rohm, \npb{237}{1984}{553.}}
\lref\ros{M.A.R. Osorio, \ijmpa{7}{1992}{4275.}}
\lref\rsdual{E. Witten, \npb{443}{1995}{85.}}
\lref\rpolc{J. Polchinski, \prl{75}{1995}{4724.}}
\lref\rdir{M.B. Green, \plb{266}{1991}{325.}}
\lref\rseipol{J. Dai, R.G. Leigh and J. Polchinski, \mpla
{4}{1989}{2073\semi}
R.G. Leigh, \mpla{4}{1989}{2767.}}
\lref\rkap{J.I. Kapusta, {\it Finite-Temperature Field Theory}, Cambrige 1989.}
\lref\rcaipol{J. Polchinski and Y. Cai, \npb{296}{1988}{91.}}
\lref\rwguo{A. Erdelyi, {\it Higher Transcendental Functions}, McGraw-Hill,
New York 1953.}
\lref\rdoljack{L. Dolan and R. Jackiw, \prd{9}{1974}{3320.}}
\lref\raw{J. Atick and E. Witten, \npb{310}{1988}{291.}}
\lref\rov{M.A.R. Osorio and M.A. V\'azquez-Mozo, \plb{280}{1992}{21} (\bb{9201044});
\prd{47}{1993}{3411}(\bb{9207002}).}
\lref\rshifmanb{M. Shifman, Ed., {\it Instantons in Gauge Theories} 
(World Scientific, Singapore, 1994) p. 237--246.}   
\lref\rhama{K. Hamada, {\it Vertex operators for super-Yang-Mills
and multi-D-branes in Green--Schwarz superstring}, KEK-TH-504 
(\bb{9612234}).}   
\lref\rgge{M.B. Green and M. Gutperle, {\it Configurations of
two D-instantons}, DAMTP-96-110 (\bb{9612127}).} 
\lref\rgut{M. Gutperle, \npb {444}{1995}{487} (\bb{9502106}).}
\lref\rvet{M. Laucelli Meana, M.A.R. Osorio and J. Puente
Pe\~nalba, {\it The string density of states from the
convolution theorem}, FFUOV-97/01 (\bb{9701122}).}
\lref\rbarcua{J.L.F. Barb\'on, {\it Remarks on the Classical Size of
D-Branes}, CERN-TH/97-52 (\bb{9703138}).}   


\line{\hfill CERN-TH/96-361}
\line{\hfill IASSNS-96/127}
\line{\hfill EHU-FT/9701}  
\line{\hfill {\tt hep-th/9701142}}
\vskip 0.2cm

\Title{\vbox{\baselineskip 12pt\hbox{}
 }}
{\vbox {\centerline{Dilute D-Instantons at Finite Temperature  }
}}

\centerline{$\quad$ {\caps J. L. F. Barb\'on
 }}
\smallskip

\centerline{{\sl Theory Division  CERN}}
\centerline{{\sl CH-1211 Geneva 23, Switzerland}}
\centerline{{\tt barbon@mail.cern.ch}}

\vskip 0.2in
\centerline{$\quad$ {\caps M. A. V\'azquez-Mozo}}
\centerline{{\sl Institute for Advanced Study}}
\centerline{{\sl Princeton, NJ 08540, U.S.A.}}
\centerline{{\sl and}}
\centerline{{\sl Dept. de F\'{\i}sica Te\'orica\foot{Address after
January 1st, 1997.}}}
\centerline{{\sl Universidad del Pa\'{\i}s Vasco}}
\centerline{{\sl Apdo. 644, E-48080 Bilbao, Spain}}
\centerline{{\tt wtbvamom@lg.ehu.es}}

 \vskip 0.2 in

We discuss Dirichlet instanton effects on type-IIB string 
thermodynamics. We review some  general properties of  dilute
D-instanton gases  and use the low-energy supergravity solutions
to define the normalization of the instanton measure, 
as well as the effects of long-range interactions.
Thermal singularities in the single-instanton sector are due
to tachyonic winding modes of Dirichlet open strings. Purely
bosonic D-instantons induce in this way hard infrared singularities
that ruin the weak-coupling expansion in the microcanonical
ensemble. However, type-IIB D-instantons,  give smooth contributions 
at the Hagedorn temperature, and the  
induced mass and coupling of the axion field are insufficient
to change the first-order character of the phase transition in the
mean field approximation.


\Date{01/97}


\newsec{Introduction}

Since its prehistoric days, string theory has been formulated in
a perturbative version in which the string coupling constant
is assumed to be small. However,  the shortcomings
of this approach were early realized,  
since a great deal of physically interesting situations
would involve some kind of non-perturbative phenomena. In fact,
one could say that a weakly coupled description of the physics
at the string scale needs non-perturbative input to be predictive
at all, because of the non-renormalization theorems implied by
supersymmetry, which is itself  a necessary ingredient for consistency 
of the weakly coupled description.         
In spite of
this, non-perturbative string physics has remained an elusive realm  
where no significative progress was made at a quantitative
level,  until the recent advent of
string dualities (for a review see \refs\rprev). One of the lessons 
to be learned from this second string revolution is the dynamical
relevance of generalized soliton states in string theory.   
 Since these states
have to decouple in the weak-coupling limit, their masses scale
with inverse powers of the string coupling constant. In the case
of type-II strings, string duality implies the existence of 
non-perturbative states carrying R--R charge and having masses of the 
order of $\lambda^{-1}$. It was realized by Polchinski \refs\rpolrr\
that those states are provided by D-branes \refs\rpolsem. 

On general grounds, a $p$-brane with a tension of order $1/\lambda$ 
produces non-perturbative effects of order $e^{-V_{p+1}/\lambda}$,
where $V_{p+1}$ is a characteristic Euclidean world-volume associated
semiclassically to a $(p+1)$-dimensional non-contractible cycle in
target space.
This means that, for the first time \refs\rpolins, we have an
explicit semiclassical construction of the ``stringy"
 non-perturbative
effects already expected on the basis of
 perturbation theory estimates
\refs\rshenker.      The   D-brane description provides a 
perturbative treatment of the collective coordinate dynamics,
something very problematic in other approaches to soliton
quantization in string theory. It is then very interesting
to understand the weak-coupling quantization of such objects,
non only for the applications to duality, but more generally
because they apply  to more  realistic
 situations without
extended supersymmetry.

Among different $p$-dimensional D-branes, D-instantons ($p=-1$) 
are special for several reasons. In spite of their stringy
origin, they appear to be point-like objects
when probed with closed strings at high energies, contrary to 
higher-dimensional D-branes, which have an effective size of order 
$\sqrt{\alpha^{'}}$ \refs\rpoint. Another important point is the  
 fact that  the corresponding $e^{-1/\lambda}$ effects are present
in the ten-dimensional type-IIB theory while all the rest of
Euclidean D-brane instantons are suppressed in the decompactification
limit. As a result, they represent the technically simplest case.    
At a dynamical level, D-instantons differ from more conventional
Yang--Mills instantons in that they do not have size moduli and,
more importantly, there are long-range Coulomb interactions between
instantons and anti-instantons, in complete analogy with the 
well-known monopole instantons producing a mass gap in 2+1 gauge
theories \refs\rpoly.   

 In the present paper we will
be concerned with the physical consequences of D-instanton effects
in the thermodynamics of ten-dimensional type-IIB  strings\foot{
Effects of Dirichlet boundaries on thermal strings were considered
in the past by M. Green with a  different motivation
\refs\rgreent.}.
String theory at finite temperature \refs\raorw\ is one scenario in which 
non-perturbative effects have frequently been  invoked to account
for  a number 
of 
perturbative riddles. Probably the most interesting of these is 
the existence of a critical  temperature (the so-called Hagedorn
temperature) at which the canonical description of the string collective 
breaks down. In switching from the canonical to the microcanonical
ensemble a phase of negative specific heat appears above the critical
temperature, indicating an instability of the string gas\foot{We
are describing here the generic situation. If all spatial dimensions are
compactified the negative specific heat phase disappears and the
Hagedorn temperature becomes a maximum temperature both in the 
canonical and microcanonical ensembles.}.
This Hagedorn ``crisis" has been the subject, over the years,
 of much controversy
 as to whether it really signals a phase transition into a, 
yet mysterious, high-temperature phase of string theory or it just marks 
a maximum temperature of the string gas (for a recent entry, see 
\refs\rvet).  

Generally speaking, instanton effects can dominate at weak coupling
only when considering quantities without any perturbative contributions.
Since this is not the case for the usual thermodynamical functions,  
such as the canonical free energy or the microcanonical entropy function, 
we do not expect important modifications of the perturbative lore at
a qualitative level. In this sense, we discuss instanton-driven
thermal singularities in the corresponding Dirichlet string diagrams
as a check of the consistency of the D-instanton weak-coupling
expansion.    

Within the canonical approach, we find that the possible instanton
singularities are hard but located well inside the Hagedorn domain, in
a region of temperatures where the string diagram expansion does not
make much sense.  
We have also  studied the microcanonical description of
the string gas near the Hagedorn transition, including D-instanton 
contributions to the density of states, computed as an inverse Laplace
transform of the analytically continued canonical partition function. In
this case the temperature singularities beyond the Hagedorn domain do
have a concrete physical effect, as subleading terms in a high-energy
expansion.
 Our results show that whenever new instanton-driven singularities
appear, as in the case of purely bosonic strings,
 the whole  instanton expansion
makes no sense at all, being even more singular than  could have
been expected on the basis of the zero-temperature instabilities of bosonic
strings. This instability is characterized by uncontrollable power corrections
in the high-energy asymptotics of the microcanonical density of states. 
On the other hand, for type-IIB strings in ten dimensions,
with a well-behaved perturbation theory, we find that D-instantons
do not induce new singularities, and the Hagedorn phase transition
seems to be triggered by the same tachyonic winding mode as in
standard perturbation theory.

   A genuine instantonic effect in the type-IIB theory is the
generation of a non-perturbative mass and non-derivative couplings
for the axion scalar field, protected in perturbation theory by
a Peccei--Quinn symmetry. This is a situation where instantons 
can in principle produce qualitatively important effects through the
axion dynamics. We have re-examined the role of the axion in the
Atick--Witten 
mean field theory analysis of the Hagedorn phase transition  
 and found that, while D-instantons tend to induce a 
second-order transition, the dynamics is still
 dominated by the dilaton,
which produces a first-order phase transition with a critical 
temperature smaller than the Hagedorn temperature.    

We will not consider in this paper the case of type-IIA strings.
Due to the presence of the thermal non-contractible circle, Euclidean
world-lines of Dirichlet 0-branes produce Wilson--Polyakov loop 
contributions of order
$e^{-C \beta /\sqrt{\alpha'} \lambda} $, which are interesting
to study (see \refs\rgbu\ for  work in this direction). 
The T-duality that
should relate the IIA and IIB cases works differently with  
thermal boundary conditions \refs\rdual. In any case, temperature
T-duality (also called $\beta$-duality) is probably unphysical and should 
be broken.    

The plan of the paper is as follows: we begin Section 2 by reviewing
some general facts about D-instantons and the construction of ref. \refs\rggp\
of D-instanton solutions in type-IIB supergravity. After this we apply the 
collective coordinates method \rtofito\ to obtain 
 the instanton measure and compute
the effective action for the massless scalars (axion and dilaton)  
induced by the long-range Coulomb interactions.
 In Section 3 we start our discussion of
D-instanton effects in string thermodynamics by studying the D-instanton-induced
singularities in the canonical free energy and its relevance in the 
microcanical description of strings near the Hagedorn temperature. Section 4 
will be devoted to the 
study of the relevance of D-instantons in the low-energy effective theory
description of the Hagedorn transition. Finally, in Section 5 we will
 summarize
our conclusions. A discussion of fermionic collective coordinates
is included in Appendix A.  For the sake of completeness,
 we will discuss the computation of the
microcanonical density of states in perturbation theory in Appendix B,  
while in Appendix C we
 will review the construction of higher-order amplitudes
in open bosonic string theory.

\newsec{Dirichlet Instanton generalities}

Formally speaking, D-instantons correspond to the ``degenerate'' case of
$p=-1$ D-branes, i.e. the world-volume is a point in space-time
and the collective dynamics reduces to a finite-dimensional integral
over moduli space. The action and interactions are determined by a
theory of open strings without translational zero modes, and  
it was originally obtained as a T-dual of standard open strings with
respect to all translational isometries.

A complete perturbative expansion for a dilute D-instanton gas was
obtained in \refs\rgreengas, \refs\rpolins\
 on the basis of T-duality and cluster 
decomposition.
The complete partition function takes the form
\eqn\partfunc
{{\cal Z} = \sum_{n_+,n_- = 0}^{\infty} {1\over n_+ ! n_- !}
\prod_{j=1}^{n_+}\int d\mu^+_j  \prod_{k=1}^{n_-} \int 
d\mu^-_k
 \, e^{-S_{(n_+,n_-)}}, }
where $d\mu^{\pm}$ denotes the collective coordinates measure for a single
D-instanton and   the action in the $(n_+, n_-)$ instanton sector is
\eqn\action
{S_{(n_+,n_-)} = \Gamma_0 + \sum_j \Gamma_j + \sum_k \Gamma_k +
\sum_{(j_1, j_2)} \Gamma_{(j_1,j_2)} +\sum_{(k_1,k_2)}
\Gamma_{(k_1,k_2)} + \sum_{(j,k)} \Gamma_{(j,k)} + {\rm 3\,\,body 
} ;} 
here the index $j$ refers to instantons and $k$ to anti-instantons.
We have an expansion in irreducible many-body interaction terms, each
of them given by the sum of connected string diagrams with a number of
boundaries attached to instantons, anti-instantons, or both.
Specifically,
\eqn\genterm
{\Gamma_{(j_1,\cdots;k_1, \cdots)} = \sum_{g=0}^{\infty}
\sum_{N^+_1,\ldots=0}^{\infty} \sum_{N^-_1, \ldots =0}^{\infty}
{\lambda^{2g+\sum N^+ + \sum N^- -2} \over N^+_1 ! \cdots N^-_1 !
\cdots } W(g, N^+_j, N^-_k) .}
Here $N^{\pm}$ denote the numbers of boundaries attached to the same
instanton or anti-instanton, and $\lambda$ stands for 
the string coupling constant. The first term $\Gamma_0$ is the standard
perturbative sum of string diagrams {\it in vacuo}, and   
 the bare
instanton action is given by
$
\Gamma_{\pm} = {|Q|\over \lambda}$,  
where
  $|Q|= W_{0,1}$ is numerically equal to the disk amplitude with Dirichlet
boundary conditions. Diagrams without handles   correspond
to classical interactions  between the instantons, the leading one
coming from the cylinder diagram, which corresponds at low energies
to the interactions induced by the overlap of the instanton tails at
long distances (see Appendix A for an explicit expression).
  The expansion \genterm\
and \partfunc\ provides a complete perturbative treatment of the
instanton interactions, including the purely classical ones,
giving a stringy version of a perturbative constrained instanton
expansion \refs\raffleck, \refs\rshifmanb.  

    In an operator formalism we can translate the Dirichlet boundary 
conditions on the world-sheet fields into boundary states constructed
as convenient coherent states in the single-string Hilbert space. The
simplest description of such states in the supersymmetric case is
achieved in the light-cone frame, using the Green--Schwarz formalism.  
In the notation and conventions of \refs\rgreenf,  the bosonic
(anti-) D-instanton boundary state is a solution of the constraints
$(\alpha^i_n - \balpha^i_{-n} )|I_{\pm}, p\ket =0$, $(S^a_n \pm i           
\bS^a_{-n} ) |I_{\pm}, p\ket =0$ and may be written as the coherent
state   
\eqn\bs
{|I_{\pm}, p\ket = {\rm exp}\sum_{n=1}^{\infty}\left( {\alpha^i_{-n}
\balpha^i_{-n}\over n}
 \mp i S^a_{-n} \bS^a_{-n} \right) \, |0_{\pm}, p\ket ,}
where $S^a$ are Green--Schwarz fermions, transforming in the ${\bf 8_s}
$ of $SO(8)$, the transverse rotation group. The ground states in
\bs\ are  
the standard massless scalars  of the type-IIB string
$
|0_{\pm}, p\ket = {1\over 4}\left(|p\ket |i\ket |{\overline i}\ket
\mp i |p\ket |{\dot a}\ket |{\overline {\dot a}} \ket \right)$ 
 satisfying\foot{As usual, the indices $i,a, {\dot a}$ run in the
${\bf 8_v}, {\bf 8_s }$ and ${\bf 8_c}$ of $SO(8)$, respectively.}
 $\bra 0_{\pm}, p| = (|0_{\mp},p\ket)^{\dagger}$, $\bra
0_{+}, p |0_{+}, p' \ket = 0$, $\bra 0_{+}, p | 0_{-}, p' \ket =
\delta^{10} (p+p')  $, and the position space vacuum is defined
by the  standard Fourier transformation $|x_0 \ket = \int {d^{10} p 
\over (2\pi)^5} |p\ket e^{ip x_{0}}$.

These boundary states break half of the $32$ real supersymmetries of
the type-IIB theory in ten dimensions. Writing the charges in the
Green--Schwarz formulation as $Q^a$ and $Q^{\dot a}$, and similarly 
for world-sheet right-movers, we have ${1\over \sqrt{2} }(Q \pm i
{\overline Q}) |I_{\pm},p \ket =0$. Therefore, the ten-dimensional
D-instantons have a total of $16$ fermionic zero modes. Because
of this BPS character, Dirichlet amplitudes with either instantons
or anti-instantons, but not both, have unbroken supersymmetries 
that can be used to prove vanishing theorems by means of world-sheet
Ward identities (see Section 3.2). This is not the case for amplitudes 
involving {\it both} instantons and anti-instantons, or for any amplitude 
at finite
temperature, since the supersymmetry is then completely broken by
the vacuum boundary conditions.  

At finite temperature, the boundary states are exactly given by
\bs\ in the oscillator sectors. At the level of the zero modes we
simply have to implement the Kaluza--Klein quantization of the
time-like momenta in the closed string channel $p^0 =2\pi n / \beta$,
and the completeness relation for discrete thermal momenta becomes
the usual $\bra n |m\ket = \beta^{-1}\delta_{n,m}$. In the open string
channel, as we shall see in more detail below, these discrete states
are interpreted as winding modes of open strings around the thermal
circle, which now carry a winding topological charge because their
ends are constrained to lie at the instanton location.

The extreme dilute approximation is defined by dropping all interaction
terms in \genterm\ beyond the single-instanton effective action. For
example, the contribution to the free energy or vacuum energy in a
toroidal space ${\bf R}^9 \times S^1_{\beta}$ is given by
\eqn\dilute
{e^{-\beta F(\beta) } = e^{-\Gamma_0} \sum_{n_+ = n_-}{1\over n_+ ! n_- !}
 \left( \int
d\mu_+ e^{-\Gamma_1^+} \right)^{n_+} \left( \int d\mu_- e^{-\Gamma_1^-}
\right)^{n_-}, }
where the constraint of neutrality $n_+ =n_-$ appears because the
D-instantons are sources of R--R flux, and charge neutrality is necessary
for consistency in a compact space. 
Assuming $d\mu_+ = d\mu_-$ and writing the constraint as
 $\delta_{n_{+},n_{-}} =
\int_{0}^{2\pi} {d\theta \over 2\pi} e^{-i\theta (n_{+}-n_{-})}$, we can 
exponentiate the result to
\eqn\dillu{
e^{-\beta F(\beta)} = e^{-\Gamma_0} \int_{0}^{2\pi} {d\theta\over 2\pi}
{\rm exp}\left( 2 \int d\mu\, e^{-\Gamma_1 }\,{\rm cos}\,\theta \right). }  

If the measure $d\mu$ contains
 fermionic collective coordinates due to fermionic
zero modes of the instantons, the non-perturbative corrections to the
free energy in \dillu\  
 clearly vanish.
At finite temperature, however, all supersymmetries are broken by the
vacuum boundary conditions (thermal antiperiodicity of fermions). We
take the attitude that collective coordinates should be associated only
to symmetries broken by the localized instanton, but restored asymptotically
at infinity; otherwise we have a spontaneously broken symmetry and the
collective modes are really Goldstone bosons. So, at finite temperature
we expect the integral over the moduli to produce simply a factor of
the ten-dimensional volume. The sum over vacuum diagrams $\Gamma_0$ is
just the perturbative free energy, so we end up with a $\theta$-band  
\eqn\freedil{
F(\beta, \theta)_{\rm dilute} 
 = F(\beta)_{\rm pert} -C\cdot{\rm Vol}_9 \cdot
  e^{-\Gamma_1 (\beta)} \,
 \cos \theta,}
with $C$ some positive
 constant of order ${\cal O}(1)$ in string
units. In the next section we will see that $\theta$ is just the zero mode of
the dynamical axion field, so that we have to minimize \freedil\ and
relax to $\theta=0$. It is already clear from eq. \freedil\ that
hard singularities in the one-instanton effective action $\Gamma_1 (
\beta)$ could in principle change the critical behaviour of the
free energy. For example, a power behaviour at the Hagedorn 
temperature $\Gamma_1 (\beta)_{\rm sing} \sim - |\beta-\beta_H|^{
\alpha}$, with $\alpha \ge 0$, would turn the Hagedorn temperature
into a maximum temperature, although a safer interpretation would
be that the instanton expansion (in the sense of a weak-coupling
expansion) breaks down badly, because a
term nominally suppressed as $e^{-|Q|/\lambda}$ would dominate the
physics.       

The dilute approximation is constructed as a perturbative expansion
around the gas of free instantons, according to the general expressions
\partfunc --\genterm. 
  The first  corrections come from the 2-body interaction terms $\Gamma_{j,j'}$,
$\Gamma_{k,k'}$ and $\Gamma_{j,k}$, which at low energies reduce to
Coulomb interactions. We shall see in the next section that these terms
generate periodic potentials for the carriers of these forces.

\subsec{Low-Energy Solutions}

At the massless level the D-instanton couples to the dilaton--graviton
system and the R--R axion field. It is therefore possible to describe the
D-instanton projection over the massless fields as a classical solution
of the corresponding type-IIB supergravity theory. In this section we
review some properties of the solution described in \refs\rggp\ and
use it to study some dynamical aspects, which are most easily handled in
the field theory framework, such as a derivation of an effective 
Lagrangian by integrating out the instantons in a dilute approximation.

In order to match the string coupling dependence with the previous
stringy description of the D-instanton gas, we need to use a metric
frame which approaches the string frame at least asymptotically. At
the same time we would like to have a space-time independent Newton
constant by decoupling the dilaton from the trace of the gravitational
metric. We can reconcile these two requirements by going to the
``modified Einstein frame"  \refs\rmalda,
  which is related to the string frame
by the Weyl rescaling
\eqn\wres{
(g_{\mu\nu})_{\rm str} = e^{ {\phi-\phi_{\infty} \over 2}} g_{\mu\nu},}
which differs from the standard Einstein frame by a constant rescaling
by the asymptotic value of the string coupling $\lambda = \lambda_{\infty}
\equiv e^{\phi_{\infty}}$. In the following we will use this normalization
of the metric, which leads to a slight change with respect to 
 the string coupling
conventions of \refs\rggp.

Since the D-instanton is magnetically charged with respect to the
 R--R 8-form
potential of the type-IIB theory, it arises as a classical solution of
the Euclidean action 
\eqn\eqac{ S= {1\over 2\kappa^2_0 \lambda^2} \int 
\sqrt{g} \left(
-R + {1\over 2} (\partial \phi)^2 + {\lambda^4 \over 2} e^{-2\phi} 
F_9^2 + ...\right). }
Here $\kappa_0^2 \sim \alpha'^4$ relates the string scale and the
Newton constant\foot{There is a natural normalization of this parameter
in the type-IIB theory as the  appropriate ratio of the D-string and
fundamental string tensions, see \refs\rpolrev.}.
For a single D-instanton located at the origin we have
\eqn\instsol{\eqalign{
ds^2_{c\ell} =& dr^2 + r^2 d\Omega^2_9 \cr 
e^{\phi_{c\ell}} =& \lambda \left( 1 + {\lambda \over 8} 
{|C|\over r^8}\right)
\cr 
(F_9)_{c\ell} =& C \, d\Omega_9 , }}
with $C$  a constant that will be determined shortly. The most important
property of \instsol\ is the total classical decoupling of gravity in
the Einstein frame (or modified Einstein frame here), as the classical
metric is flat. This fact makes the low-energy analysis of these instantons
mostly independent of the more complicated gravitational
sector.   

The solution \instsol\ is ``self-dual" in the sense that
\eqn\bps{
d\phi_{c\ell} + \lambda^2 e^{-\phi_{c\ell}} * (F_9)_{c\ell} =0,}
which is in fact the BPS condition for preserving half of the total
$N=2$ ten-dimensional type-IIB supersymmetry. As stated before, this
property means that multi-instantons are easily constructed by
direct superposition of the single-instanton solution in \instsol.

The local string coupling field $\lambda (x) = e^{\phi (x)}$ is harmonic
in bulk, $\pt^2 e^{\phi_{c\ell}} =0$, from which we derive the useful
identity $(\pt \phi_{c\ell} )^2 = -\pt^2 \phi_{c\ell}$, which is
 valid away from the
instanton location.
Taken together with the BPS property \bps, it allows a simple calculation
of the instanton action with the result
\eqn\inac{
S_{c\ell} = {1\over 2\kappa_0^2 \lambda^2} \int \left( 
\half (\pt \phi_{c\ell} )^2
+ {\lambda^4 \over 2} (F_9)^2_{c\ell}\right)
 = {1\over 2\kappa^2_0 \lambda^2} \cdot
\lambda |S^9| |C| = {|Q| \over \lambda},}
with $|S^9|$ the volume of the unit nine-sphere and $|Q| \equiv W_0$ the
disk amplitude. This fixes the constant $C= \pm
{2\kappa_0^2 |Q| \over |S^9|}$.
The sign of $Q$ is defined as  the sign of the flux 
\eqn\flus{ {1\over 2\kappa^2_0 |S^9|} \oint_{\infty} F_9 = Q, }
and it is positive or negative 
according to the instanton or anti-instanton character of the solution. 

We have seen that the instanton is a Coulomb ``magnetic" source for
$F_9$, much in the same way as 't Hooft--Polyakov monopoles arise
as instantons in 2+1 models. By Dirac duality, the D-instanton must
be electrically charged with respect to the axion field dual to the
8-form. In the Euclidean formalism we must implement the Dirac duality
 by means of a Gaussian transformation of the 
$A_8$ path integral by the change of variables
(see \refs\rmio\ for a general treatment of such manipulations)   
\eqn\cha{[\CD A_8]_{\rm inst}
 \rightarrow \CD F_9\,\, \delta [dF_9 -2\kappa_0^2 *J_0 ] 
\sim \int \CD a\, \CD F_9 \,e^{-i \int a\wedge \left( {1 \over
 2\kappa_0^2} F_{9}
- *J_0 \right)},}
which introduces the axion field as a Lagrange multiplier. In changing
variables from $A_8$ to the field strength defined locally as
 $F_9 \sim dA_8$, we
have to  exhibit explicitly the magnetic sources $J_0$, which prevent
$F_9$ from being a closed form. Under this change of variables we
can establish the duality by direct evaluation of the $F_9$ path
integral, leading to
\eqn\du{\int [\CD A_8 ]_{\rm inst} \,e^{-{\lambda^2 \over 4\kappa_0^2} 
\int e^{-2\phi} |F_9 |^2 } \sim \int \CD a \,e^{-{1\over 4\kappa_0^2 \lambda^2}
\int e^{2\phi} |da|^2 - i\int a\wedge *J_0}}
and the saddle point in $F_9$ determines the Euclidean duality
\eqn\duu{ (F_9)_{\rm saddle} = i\, {e^{2\phi}\over \lambda^2}  * da.  }
Using \duu\ we can infer the classical axion profile in the instanton
background that would solve the axionic action
\eqn\axa{ S_{\rm axionic} = {1\over 2\kappa_0^2 \lambda^2} \int 
\sqrt{g} 
\left( -R + \half (\pt \phi)^2 + \half e^{2\phi} (\pt a)^2 \right) + {\rm
 sources} }
and it is given by \eqn\bala{da_{c\ell} =
 i\, e^{-\phi_{c\ell}} \,d\phi_{c\ell},} from
which we can integrate the axion field as
\eqn\inte{a_{c\ell} - a_{\infty} = -i \left( e^{-\phi_{c\ell}} - {1\over
\lambda}\right). }
Notice that a new integration constant appears, the asymptotic value
of the axion at infinity, and that the local axion field is complex. As
a result, the full instanton action comes from the source term, as
the bulk contribution from \axa\ vanishes on account of the complex
axion field. The  normalization for the ``current" of a D-instanton
located at $x_0$ is adjusted as  
\eqn\inden{
J_0 = Q \, \delta^{10} (x-x_0),  
}
so that it leads to the correct statistical weight 
\eqn\wei{
e^{-S_{c\ell}} = e^{-S_{\rm sources}} =
 e^{-i\int a_{c\ell} \wedge *J_0 } = e^{-i a_{c\ell} (x_0) Q}
= e^{-{|Q|\over \lambda} -iQa_{\infty}}.}
In view of the $SL(2,{\bf R})$ duality of type-IIB supergravity,  
  the natural  prescription is to take  
$a_{\infty}$ as a real number, which is in turn interpreted as a theta
parameter, effectively compactified as an angular variable of period
$2\pi /Q$ when the non-perturbative dynamics breaks the continuous duality
symmetry to $SL(2,{\bf Z})$.
 
Finite-temperature solutions are straightforward in terms of the
single-instanton solution \instsol. Because of the BPS property
we are free to superimpose multi-instantons, and a periodic configuration
in Euclidean time is easily obtained by summing all discrete translations
of period $\beta$ in the time direction. For example, we have for
the dilaton profile
\eqn\solfint{
e^{\phi_{c\ell}} = \lambda + {\lambda^2 \over 8} 
 \sum_{n\in {\bf Z}} {|C|\over |x-x_0 + n\beta
e^0 |^8}, } 
with $e^0$ a unit vector in the time direction. Now the action has to
be integrated only over the cylinder $0 < x < \beta$ in the time
direction, and the result is easily checked to be the same as in
the vacuum case $S_{c\ell} = |Q|/\lambda + ia_{\infty} Q$. In fact,
when computed in the axionic  form \axa\ and \wei, the result is
trivial,  
 because it comes from a delta-function-localized source
term. This  is compatible with the full stringy determination,
where the leading classical action comes from the disk diagram. Since
the disk boundary is mapped to the instanton location, no winding modes
are allowed and no temperature dependence can appear to this order. We 
also see that $a_{\infty} Q$ is nothing but the theta parameter introduced
in \dillu\ to enforce charge neutrality in the presence of compact directions,
and therefore $a_{\infty}$ is to be integrated between $0$ and $2\pi/Q$.

\subsec{Instanton Measure}

The instanton measure is generated by quantum corrections to the
classical solution, and it naturally splits into two qualitatively
different contributions: zero modes, inducing the collective coordinates
measure $d\mu$, and perturbative quantum corrections due to non-zero
modes, which are summarized in the non-classical part of the one-instanton
effective action 
\eqn\effacun{
\Gamma_1 ({\rm quantum}) = \sum_{g>0} \sum_{N>0} {\lambda^{2g+N-2} \over
N!} W_{g,N}, }
 which contains string diagrams with at least one loop of closed strings,
and at least one instanton insertion via a Dirichlet boundary. On the
other hand, the ``classical" action is given by the genus-zero terms
\eqn\effaclas{ \Gamma_1 ({\rm classical}) = S_{c\ell} = 
{W_{0,1} \over \lambda} +
\sum_{N>1} {\lambda^{N-2} \over N!} W_{0,N}.}
In type-IIB strings, space-time supersymmetry together with the BPS character
of the D-instantons implies that $W_{0,N} =0$ for $N>1$ and therefore there
are no perturbative corrections to the disk contribution to the classical
action. This is not the case for purely bosonic D-instantons. 

The full  one-loop  measure is given by 
 \eqn\unme{d\mu \,
 A_{1-{\rm loop}} = 
d\mu \,{\rm exp}\left(- \sum_{N>0} {\lambda^N \over N!} W_{1,N} \right)} in
the full string theory, which should go over the standard determinant
factor in the low-energy $\alpha' \rightarrow 0$ limit
\eqn\lowen{
A_{1-{\rm loop}} = 1 - \lambda W_{1,1} + \CO (\lambda^2) \rightarrow 
\left( {\rm det}' S^{(2)} ({\rm inst})_{\rm g.f.} \over {\rm det}' S^{(2)} 
 ({\rm vac})_{\rm g.f.}
\right)^{-\half} \, \left( {{\rm det}' D ({\rm inst}) \over {\rm det}' D
({\rm vac})}\right). }  
Here $S^{(2)}_{\rm g.f.}$ stands for the bosonic
 fluctuation kernel of the 
low-energy theory with possible gauge-fixing (ghost) components, and
$D$ is the corresponding fermionic kernel.   
There are a few subtleties in the limit \lowen: each determinant
factor in the right-hand side contains an infinite series of
powers of the string coupling. In the stringy construction, the 
states running in the propagators in \effacun\ are just free string
states in flat space;  therefore the real loop-counting parameter
of the low-energy theory is just the number of closed string loops.
If we would rearrange \effacun\ into string field theory diagrams
and take the low-energy limit we would find for example the fermionic
determinant ratio in the form
\eqn\resum
{\sum_{n>0} {(-\lambda)^n \over n} \left( {1\over \gamma\cdot\pt} 
\varphi_{c\ell} \right)^n = -{\rm Tr}\,{\rm log}\,\left( 1+ {1\over
\gamma\cdot \pt} \lambda \varphi_{c\ell} \right), } 
where $\varphi_{c\ell}$ represents the massless bosonic profiles
of the instanton coupling to fermion bilinears in the low-energy
effective theory. In the full string theory $\varphi_{c\ell}$ is
associated to the tadpole $\bra V_{\varphi} | I \ket$ in the 
D-instanton background. In fact, the existence of non-vanishing 
tadpoles more or
less implies that this is the right interpretation of the low-energy
limit.  
Notice also that, 
 in the low-energy description,  the instanton
fields are not quantitatively significant beyond a string-scale ``core",
and therefore we may perturbatively expand the profile function. Since
all profiles are controlled by the dilaton profile, we have an effective
expansion in powers of the local 
string coupling. This is explicitly seen
in the form of the only non-trivial one-loop kernel, from the axion--dilaton
sector: 
\eqn\axdilk{
\CK_{\phi, a} = \left(\matrix{-\pt^{2}-2(\partial\phi_{c\ell})^{2} &
2i  \partial_{\mu}\left(e^{\phi_{c\ell}}\right)\partial^{\mu} \cr 
-2i \partial_{\mu}\left(e^{\phi_{c\ell}}\right)
\partial^{\mu}& -\partial_{\mu} e^{2\phi_{
c\ell}}\partial^{\mu}}\right).}
In practice, it is somewhat complicated to write explicit expressions
for the right-hand side of \lowen, due to the presence of self-dual
four-forms in the type-IIB theory, but there is no problem in giving
a formal definition of the stringy expression \effacun. We will not
need an explicit expression for the non-zero mode quantum effective
action, which we denote by $A$ in the remainder of the paper. We just
point out the general structure $A\sim 1+ {\cal O}(\lambda)
$, which is rather
obvious in the stringy construction of $\Gamma_1$, as in \effacun.

This simple situation must be modified when quasi-zero modes appear
for some values of the parameters. A characteristic example is the
low-temperature limit $\beta\rightarrow\infty$ of the fermionic
determinants, which develop a fermionic zero mode as supersymmetry
is restored in the zero-temperature limit. Such quasi-zero modes
must be subtracted and treated in terms of collective coordinates.

In \partfunc, the collective coordinate measure  $d\mu^{\pm}$ was not 
specified. Dirichlet instantons are distinguished from familiar 
Yang--Mills 
instantons in that they do not have scale moduli,  which just
conforms with the lack of scale invariance of the supergravity theory. In
the full stringy description it is also clear that the only data specified
by the boundary state \bs\ concern the space-time position of the instanton
$x_0$. Then, we have just ten bosonic collective coordinates for the 
position,
and a number of fermion collective coordinates associated to the zero
modes generated by broken supersymmetries; sixteen of them in the full
type-IIB theory. We will discuss here just the bosonic collective
Jacobian, because it is the only one relevant at high temperatures,  
 postponing to Appendix A a review of D-instanton 
fermionic collective coordinates, with emphasis on the treatment of
the low-temperature quasi-zero modes.   

We can easily fix the leading term of the collective coordinate measure
in the low-energy supergravity theory. Since the low-energy solution
\instsol\ in the Einstein frame is only non-trivial along the axion--dilaton
components in field space, we consider the zero-modes of the corresponding
fluctuation operator $\CK_{\phi,a}$ written in \axdilk. The relevant
quadratic form is
\eqn\cuad{
S = S_{c\ell} + \half \bra (\phi_q, a_q ) |\CK |(\phi_q, a_q )\ket + ...}
The inner product defining the path integral measure for the quantum
fluctuations $\phi_q, a_q$ is given by  
\eqn\inn{ \bra(\phi_q, a_q)|(\phi'_q , a'_q)\ket = \int d^{10} \xi \left( 
{1\over \lambda^2} \phi_q^* (\xi) \phi_q' (\xi) + a_q^* (\xi) a_q' (\xi)
\right)  } 
and we have rescaled $x= (2\kappa_0^2 )^{4/5} \, \xi$ to dimensionless
variables. Then, a set of finite-norm zero-modes of $\CK$ is given
by
\eqn\zmod{ V_{\alpha} = \left( {\pt \phi_{c\ell} \over \pt \xi^{\alpha}},
{\pt a_{c\ell} \over \pt \xi^{\alpha}} \right)}
and the resulting measure is calculated following the standard algorithm in
\refs\rtofito: 
\eqn\fito{ d\mu = \prod_{\alpha=1}^{10} d\xi_{\alpha} \, \left(   
{ \bra V_{\alpha} | V_{\alpha} \ket \over 2\pi} \right)^{\half}.}
Using the classical solutions, we easily calculate the Jacobian. By
symmetry all terms in the product in \zmod\ are equal, and therefore
\eqn\cuent{\bra V_{\alpha} |V_{\alpha} \ket = {1\over 10} \sum_{\sigma =1}^{
10} \bra V_{\sigma} | V_{\sigma} \ket = {1\over 10} \int {d^{10} x \over
2\kappa_0^2} \left( {1\over \lambda^2} (\pt \phi_{c\ell})^2 - (\pt a_{c\ell}
)^2 \right) = {1\over 10}
 \left( 1 + {1\over 3}\right) {|Q| \over \lambda},}
where we have used the fact that $\pt a_{c\ell}$ is purely imaginary and
we obtain the final result
\eqn\jaco{ d\mu_{\rm bosonic} \equiv d^{10} x_0 \, J_{\rm coll} = 
{d^{10} x_0 \over (2\kappa_0^2 )^{5/4}} \left( {|Q| \over 15 \pi \lambda}
\right)^5 .}
Notice the occurrence of the usual factor  $\left(\sqrt{S_{
c\ell}} \right)^{\nu}$ for $\nu$ translational zero modes.

The integral over position collective coordinates must be supplemented
with certain constraints. This is due to the existence of short-distance
singularities in the perturbative setting described so far. 
Strictly speaking, only pure multi-instanton configurations are exact
solutions but, as is well known in the field theory case, clustering
implies that we must consider approximate solutions with well separated
 instantons
and anti-instantons 
in order to have a consistent approximation. Short-distance singularities
come from the corners of the approximate multi-instanton moduli space,
corresponding to the coincidence limit of two or more instantons or
anti-instantons\foot{At the classical level, the solution \instsol\
develops local strong coupling at a distance of the order of the 
string scale, $\ell_s \sim \sqrt{\alpha'}$, when measured in the
weakly-coupled, dual type-IIB background \refs\rbarcua.}.
 For bosonic D-instantons, both types of degeneration
lead to singularities. For type-IIB D-instantons, the BPS property
implies regularity of the pure instanton (I) or pure anti-instanton (A)
degenerations.
However, the I--A degeneration is singular already at the classical level,
since the cylinder diagram gives a divergence at half the Hagedorn distance
$\beta_H = 2\pi\sqrt{2\alpha'}$ 
of the form $\Gamma_{I-A} \sim -C
 \,{\rm log}\, |\Delta x - {\beta_H \over 2}|$,
 where
$\Delta x$ is the  I--A separation \rgbu, \rgreenf, \rbsus.   
At zero temperature, fermion zero modes increase the constant $C$,
thus enhancing the singularity   
\refs\rmipaper.    

The  ambiguities in the I--A parametrization are well known already at
the field theory level, since 
these are not topologically distinct configurations, and I--A pairs
should annihilate into perturbative states. The sharp singularity
that appears in the stringy case is one of the most interesting and
pressing problems of D-instanton dynamics, and we will not discuss
the question further in this paper. In the spirit of the dilute
approach, we will assume that a ``hard core" interaction  exists  
around all D-(anti-) instantons,  
in such a way that the integral over their positions $x_{i}$ will be 
restricted to $|x_{i}-x_{j}|>\beta_{H}/2$.  

Actually, the hard core is needed even in the I--I  
sectors. The
reason is that, as $N$ D-instantons come together at sub-stringy
distances, there are $N^2$ light modes becoming massless that   
must be treated as quasi-zero modes and included into the
collective coordinates \refs\rwittb, \refs\rgge.   
Therefore, the collective
dynamics beyond the dilute approximation involves $U(N)$ matrix
dynamics in the I--I sectors, 
 and some unknown physics determining the  I--A annihilation.    
For the applications discussed in the present paper (general
structure of singularities in the single-instanton sector and
the infrared analysis of the phase transition) we believe the
dilute approximation to be appropriate. The limitations of
our treatment should,  however,  be kept in mind.  

\subsec{Instanton Interactions and Effective Lagrangians}

In analysing the physical effects of instantons in the dilute
approximation, it is common practice to summarize the short-distance
effects induced by instantons in an effective Lagrangian, which we
can imagine as the result of integrating out the instanton fluctuations.
This approach is specially appropriate for the D-instantons, since
these have a fixed string-scale effective size. In general, one just
considers \refs\rshifmanb\ 
the corresponding amputated Green functions in the instanton
background, and reinterprets them in the local limit as inducing
a set of local operators    
\eqn\logic{
\left\bra
 \prod_{\rm bosons} (-\pt^2) \Phi \prod_{\rm fermions} (\gamma\cdot \pt)
\Psi \right\ket_{\rm inst}
 \longrightarrow \sum_{\rm local \,\,ops.}\left\bra \CO_{\rm eff}(\Phi,\Psi)
\right\ket_{\rm vac}.}  
For example, in the ten-dimensional type-IIB theory, D-instantons have
sixteen fermionic zero modes, which have to be saturated in order to
have a non-vanishing expectation value in the one-instanton sector.
  Therefore, to leading order,  
we induce operators with  $n_f$ fermions, $n_b$ bosons
 and a number $n_{\pt}$
of derivatives of the form \refs\rmipaper 
\eqn\ops
{\CO_{\rm IIB} \sim  J_{\rm coll}\, A\, e^{-|Q|/\lambda}
\,\Psi^{n_{f}}\pt^{n_{\pt}}\Phi^{n_b},   
}
with\foot{Recently, a careful analysis of
an induced bosonic operator with $8$ derivatives, an $R^4$ term, has
appeared in \rgg.} $n_{\pt} + n_f /2 = 8$. 
These are high-dimension operators, which
are not very relevant in the infrared. However, in the
situation without fermion zero modes given by the finite-temperature
boundary conditions, purely bosonic lower-dimension operators can be
generated. For example,  
 the axion gets a 
mass from the one-instanton sector of the form
\eqn\axmass{
m_a^2 = J_{\rm coll}\, A \,
e^{-|Q|/\lambda} \left| \int \pt^2 a_{c\ell} \right|^2
\sim Q^2 
\, A
\, J_{\rm coll}\, e^{-|Q|/\lambda}. }
This formula comes from the 
 single-instanton contribution to the two-point 
function $\bra a^* (x) \, a(y)\ket_{\rm inst}$ after we amputate
external propagators and take the local limit.
In order to interpret formula \axmass\ we still need to argue that
the axion effective potential attains its minimun at $a=0$. This
is easily done in this problem, because  
including the two-body interactions modulates the mass terms as in
\axmass\ into periodic potentials, the reason being that the two-body
interactions are Coulombic at long distances. Factorizing the massless 
exchange channel, we find no interactions between like-instantons at
tree-level, due to the cancellation between the dilaton and axion
exchanges, and a doubled attractive channel between instantons and
anti-instantons,
\eqn\like{\eqalign{ \Gamma_{+,\pm} (x-y)    \sim & 
\,\Delta (x-y) 
[
 \bra +|{\rm axion}\ket\bra {\rm axion}|
\pm \ket - \bra + |{\rm dilaton}\ket\bra {\rm dilaton}|\pm\ket ] \cr
=& \,\Delta (x-y)\, ( \pm Q^2 - Q^2), }} 
where $\Delta (x-y) = \bra x |\left(
-\pt^{-2}\right) |y\ket$ is the Green function
of the ten-dimensional Laplacian. 
We see that, even with the Coulomb interactions involved, we do not really
have a plasma in the sense that there are no repulsions, and attractions
are doubled with respect to the standard case.
The corresponding partition function up to two-body interaction
terms can be written as
\eqn\twobod{
\CZ_{\rm  2-body} = 
 \sum_{n_+ =n_- \ge 0}\int {(dx^+)\,(dx^-) \over n_+ !n_- !}\, 
 \left( A\, J_{\rm coll}
\, e^{-S^+_{c\ell}}\right)^{n_+} \, \left(A\,J_{\rm coll}\,e^{-S^-_{c\ell}}
\right)^{n_-} \, e^{-V_{\rm int}^{\phi} - V_{\rm int}^{a}},  } 
where $S^{\pm}_{c\ell} = \pm i|Q| a_{\infty} 
 + |Q|/\lambda $ and charge neutrality
is enforced by integrating $a_{\infty}$ between $0$ and $2\pi/Q$. The  
 Coulomb interaction potentials due to axion and dilaton 
exchange are given by a sum over I--I, A--A and I--A
 pairs (we set $2\kappa_0^2 =
1$ for simplicity throughout this section):    
\eqn\poten{
\eqalign{ V_{\rm int}^{\phi} = & - Q^2 \sum_{(j,j')}^{n_+} \Delta_{jj'} 
- Q^2 \sum_{(k,k')}^{n_-} \Delta_{kk'} - Q^2 \sum_{(j,k)} \Delta_{jk} \cr
V_{\rm int}^{a} = & +Q^2 \sum_{(j,j')}^{n_+} \Delta_{jj'} + Q^2     
\sum_{(k,k')}^{n_-} \Delta_{kk'} - Q^2 \sum_{(jk)} \Delta_{jk} },} 
where we have suppressed the self-energy terms because they cancel
in the sum $V^{\phi} + V^{a}$.  
By means of a usual Gaussian trick we can express the interactions
in terms of path integrals for non-zero modes of the axion and dilaton
fields\foot{That is, the path integral measure is taken in the functional
space orthogonal to the zero modes of the Laplacian $-\pt^2$.}  
\eqn\pin{
\eqalign{e^{-V_{\rm int}^{\phi}} =& {\rm det}^{1/2} (-\pt^2) \int
\CD a' \, e^{-\half \int dx [ (\pt a')^2 -i a' (J_0^+ - J_0^- )] } \cr
e^{-V_{\rm int}^{a}} = & {\rm det}^{1/2} (-\pt^2) \int \CD \phi' \,
e^{-{1\over 2\lambda^2} \int dx [(\pt \phi')^2 +  \lambda \phi' 
(J_0^+ + J_0^-)]}}}  
with $J_0^{\pm} = \pm |Q| 
 \sum_{i=1}^{n_{\pm}} \delta^{10} (x-x_0^{\pm})$  the instanton or
anti-instanton currents as defined in    
\inden.
Now substituting back in \twobod\ and summing the series, we arrive at
the effective ten-dimensional Euclidean Lagrangian (here we benefit
from using the modified Einstein frame):  
\eqn\eflag{\CL_{\rm eff} = {1\over 2\lambda^2} (\pt \phi')^2 + \half
(\pt a' )^2 - 2 A\, J_{\rm coll} \, e^{-|Q|/\lambda} e^{-{|Q| \over 
\lambda} \phi'} \, {\rm cos} [Q(a' + a_{\infty})],  } 
which we can understand as the Gaussian approximation to an effective
Lagrangian for the full local string coupling ${\rm log}\lambda (x) =
\phi (x) \sim \log
\lambda + \phi' (x) + \CO (\phi'^2)$ and the full axion field
$a= a' + a_{\infty}$ with an effective potential\foot{Notice that
we integrate the axion zero mode, which is natural in a compact space.
However, we do not integrate over the dilaton zero mode, which determines
the asymptotic string coupling and poses a real super-selection rule. This
is reminiscent of similar issues in Liouville field theory, and is
characteristic of runaway potentials \refs\rseishe.}   
\eqn\inducedpot{
V_{\rm eff} = -2\,A\,J_{\rm coll} \, e^{-|Q|/\lambda(x)} \, {\rm cos}(Qa).}

So, we find the standard result that, when allowed by zero-mode
counting,  instantons generate a periodic
potential for the axion, but also a runaway potential towards the trivial
free theory for the dilaton, a well-known pathology of weakly  coupled  
string theory. 

At finite temperature, the effective ten-dimensional Lagrangian
described here only makes sense for $\beta \gg \sqrt{\alpha'}$ because
we are assuming a hard core of stringy size. In the discussion of the
last section we will be interested in the critical behaviour at 
temperatures of the order of the string scale. An effective Lagrangian
approach is still useful if we concentrate on the long wave-length 
dynamics in the spatial directions.

An important point regarding \inducedpot\ is the physics it contains. 
The only corrections to the free-instanton limit are given by the
tree-level Coulomb interaction, corresponding to the cylinder diagram
$W_{0,2}$ in the full string theory. In field theory, instanton
interactions induced by one-loop diagrams in the multi-instanton
terms come from diagrams with at least one handle and two boundaries,
and are therefore suppressed by extra powers of the string coupling.
Thus, \inducedpot\ is a consistent approximation within the weak
coupling expansion.  
\newsec{Singularities of the Free Energy}
                       
In perturbation theory both the canonical free energy and the internal 
energy of a gas of ten-dimensional type-IIB superstrings are finite 
quantities\foot{For the closed bosonic string this is also true once
we remove the divergence associated with the ground-state tachyon.}
at the Hagedorn temperature $T_{H}$. This is generally interpreted 
\refs\cabpar\ in the sense that the string collective undergoes some kind of 
phase transition near $T_{H}$, after which strings may no longer be the most 
adequate degrees of freedom to describe the physics at high temperatures
\refs\raw.

One legitimate question  is whether or not the kind of non-perturbative 
effects represented by D-instantons can change in any way the critical 
behaviour of the type-IIB superstring at the Hagedorn temperature. This may
not seem to be very likely at first sight, since D-instanton effects are
exponentially suppressed with respect to the perturbative quantum
corrections at
weak coupling. On the other hand one must remember that D-instantons
are related to open strings by T-duality and that 
for open strings the Hagedorn
temperature is a maximum temperature of the Universe where the 
internal energy diverges. Therefore there could be a  scenario
in which the non-perturbative contribution would drive the internal
energy to infinity when approaching the Hagedorn temperature from below.

In general, singularities will appear in the free energy or its derivatives 
with respect to the moduli fields
whenever there are extra massless states emerging at some point of 
the moduli space.
This is also true in the thermal case, where the only modulus now is the 
temperature $T$, but the situation here is even
worse since the state becoming massless at the Hagedorn temperature
turns out to be tachyonic above $T_{H}$, rendering the whole perturbative
expansion meaningless. The squared effective mass of this state
generically scales with the inverse temperature 
$\beta$ as $
\alpha^{'}m^{2}_{\rm eff}\sim (\beta^{2}/\alpha^{'})-a$,
where $a>0$; therefore the state triggering the Hagedorn transition
corresponds to a winding string. For the closed string models this
winding mode comes from the usual solitonic sector wrapping around the 
compactified Euclidean time. The picture is different in the presence of a 
D-instanton located at $x_{0}$ since we then  have open strings whose 
endpoints are stuck at this position, and this produces topological
sectors of a new kind, corresponding to winding open strings. It is 
easy to see that these are the  states producing new 
thermal singularities
in instanton diagrams.

To be more explicit, we will illustrate the relevant points in the
context of bosonic strings and D-instantons, whose boundary states
are easily obtained by dropping the fermionic dependence in \bs.
Because of the zero-temperature tachyon instability of the bosonic
string, the present discussion is only
 illustrative and an explicit disentangling
of the singularities between thermal and non-thermal is required. 
Later, we will address the case of the type-IIB superstring where 
the absence of a zero-temperature tachyon makes the discussion more
interesting from the physical point of view. 

 The first temperature-dependent 
contribution to the free energy comes from the sphere with two boundaries 
attached to the D-instanton, i.e. the annulus with Dirichlet boundary 
conditions: 
\eqn\cil{ W_{0,2} (\beta) = -{\pi^{12} (2\pi\alpha')^{13} \over 2} 
\int_{0}^{\infty} ds \, \bra I,x_0 |\,e^{-s\Delta_{\rm closed}} |I,x_0 
\ket = -{1\over 2} \int_{0}^{\infty}{dt\over t}
 \,{\rm Tr} \, e^{-t\Delta^D_{
\rm open}}, } 
where we have exhibited the modular transformation $t=2\pi^2 /s$ 
interchanging the open and closed string channels, and the normalization
is determined by the open string channel. In the closed
channel we have tree-level propagation of closed strings between
Dirichlet boundary states with world-sheet Hamiltonian
$\Delta_{\rm closed} = {\alpha' \over 2} {\vec p}^{\;2} +{2\pi^2 \alpha' n^2
\over \beta^2}  -2 + {\rm osc} $. Notice that only momentum
modes of closed strings propagate in this channel: the closed strings
cannot wind around the thermal circle because they are mapped into the
instanton location. Discarding the zero-temperature divergences, the
first thermal singularity appears at the $s\sim 0$ endpoint, which
is therefore better parametrized in the modular transformed open
channel. In the open channel we have a one-loop vacuum amplitude of
open strings whose endpoints are completely fixed at the instanton
position (i.e. they have no momentum modes), but are allowed to
wind a number of times around the thermal circle. The corresponding
world-sheet Hamiltonian is 
\eqn\wsh{ \Delta^D_{\rm open} = {\beta^2 n^2 \over 4\pi^2 \alpha'} 
-1 + {\rm oscillators}. }
Notice the balance between the zero-point Casimir energy responsible
for the open-string tachyon, and the positive stretching energy of
the open string with winding number $n$. 
In this parametrization, the amplitude \cil\ takes the form   
 ($w=e^{-t}$, and $f(w) = \prod_{m>0} (1-w^m)$):  
\eqn\d{
W_{0,2}(\beta)=
-{1\over 2}\int_{0}^{\infty}{dt\over t}
\sum_{n\in {\bf Z}}{e^{-t\Delta^{(0)}_{D,\rm open}}\over f(w)^{24}}=
-{1\over 2}\int_{0}^{\infty}{dt\over t}
\sum_{n\in {\bf Z}}{e^{-t\left(
{\beta^{2}n^{2}\over 4\pi^{2}\alpha^{'}}-1\right)}\over f(w)^{24}}
.}
From here we see that, modulo the zero-temperature tachyon instability,
$W_{0,2}(\beta)$ diverges logarithmically 
when $\beta$ approaches $\beta_{c}=2\pi\sqrt{\alpha^{'}}$ and is divergent when
$\beta<\beta_{c}$. After an analytic continuation we find that, close to the 
critical inverse temperature,  
\eqn\dive{W_{0,2} (\beta) \rightarrow
 {\rm log}(\beta-\beta_c) +{\rm subleading}.} 
Here $\beta_{c}$ corresponds to the self-dual length under
T-duality and since $\beta_{c}=\beta_{H}/2<\beta_{H}$, it lies 
well inside the region
of values of $\beta$ for which the perturbative contribution is already
divergent. From \d\ we can read the effective masses of the states and
check that the divergence at $\beta_{H}/2$ is produced by a state with
winding number equal to 1 and
oscillator number equal to 0. As a matter
of fact we have a whole collection of critical temperatures at $\beta_{n}=
\beta_{H}/2n$ triggered by the tachyonic ground state wrapping $n$ times
around the compactified Euclidean time. 

For type-IIB strings, possible tachyonic winding modes of open
strings would appear in the Neveu--Schwarz sector, which has a
negative Casimir term. The corresponding world-sheet Hamiltonian
would be $
\Delta_{\rm open}^D (NS) = {\beta^2 n^2 \over 4\pi^2 \alpha'} - {1\over 2}
+ {\rm osc}$, and the critical point lies again at $\beta_c = \pi
\sqrt{2\alpha'}$, twice the Hagedorn temperature of type-IIB strings.
We conclude that, within the canonical ensemble, the single   
D-instanton sector gives smooth contributions to the free energy
at the Hagedorn point; new singularities, although they are  
hard, lie well beyond the range of applicability of the perturbative
D-instanton construction.       
\subsec{D-instanton Singularities and the Microcanonical Ensemble}

At face value, the fact that the singularities of $W_{2,0}(\beta)$ appear
inside the Hagedorn domain seems to indicate that they have no physical 
consequence
for the thermodynamics of the string gas. As we will see immediately,   
the fundamental clue to realize that this does not have
 to be the case lies in
using the microcanonical ensemble \refs\rfra, \rcar, \rena, 
\ric\foot{We thank
M.A.R. Osorio for pointing out the possible relevance of the microcanonical
ensemble.}. 

In the microcanonical description, the
fundamental object is the density of states 
$\Omega(E)=\sum_{a}\delta(E-E_{a})$, where the sum runs over all states in
the system. All thermodynamical quantities are
obtained from derivatives of the entropy $S=\log \Omega(E)$;
for example the microcanonical
temperature $T$ and the specific heat at constant volume $c_{V}$ are 
given by
$$
T=\left({\partial S\over \partial E}\right)^{-1}, \hskip 1cm
c_{V}=-{1\over T^{2}}\left({\partial^{2}S\over \partial E^{2}}\right)^{-1}.  
$$
Needless to say, a direct exact
computation of the microcanonical density of states
in string theory is a desperate task. On the other hand the canonical 
free energy, and thus the partition function, can be easily computed at least
at one loop just by knowing the spectrum of the theory \refs\ranalogo\alos.  
It is well known that there is a simple relation between the two quantities; 
by using an integral 
representation of the delta function in the definition of $\Omega(E)$, 
it is possible to rewrite it in terms of the 
analytic continuation of the canonical partition function 
${\cal Z}(\beta)=\exp{(-\beta F(\beta))}$ 
by the complex inversion formula
\eqn\lapinv{
\Omega(E)=\int_{c-i\infty}^{c+i\infty}{d\beta\over 2\pi i} 
\,{\cal Z}(\beta)\,e^{\beta E}
,}
where $c\in {\bf R}$ and is such that $c>{\rm Re\,}\beta_{i}$ for all the
singular points $\beta_{i}$ of ${\cal Z}(\beta)$.

Although \lapinv\ provides us with a simple way to obtain $\Omega(E)$ from
known quantities, an actual exact computation of the integral seems to
be out of the question. Instead, the authors of ref. \refs\ric\ designed a
method to get the terms of the asymptotic expansion of $\Omega(E)$
for large $E$; by pushing $c$ to the left in the complex plane
and deforming the contour to avoid the singular points, one finds that 
the leading contributions to
\lapinv\ come from the integration around the singularities of 
${\cal Z}(\beta)$;
the further 
to the left 
we place $c$  the more singular points we have to integrate around
and the more terms we get in the asymptotic expansion.
So the lesson to be learned is that {\it all}
the singularities of ${\cal Z}(\beta)$ in the complex $\beta$ plane contribute 
to $\Omega(E)$, even those lying to the left of $\beta_{H}$. 
In Appendix B we have outlined the calculation to obtain the perturbative
contribution to $\Omega(E)$ at one loop for the bosonic string,  
using the procedure just described.

On general grounds, a singularity ${\cal Z}(\beta)\sim(\beta-\beta_{c})
^{\alpha}$ leads, after an inverse Laplace transform, to a contribution
to the  asymptotic
density of states of the form 
$E^{-(1+\alpha)}e^{\beta_{c}E}$ (see Appendix B). 
A consequence of this is that the leading
behaviour of $\Omega(E)$ in the large energy regime
associated with each singular temperature will be 
given by the lowest critical exponents of ${\cal Z}(\beta)$. 
Since all the singularities have an infrared interpretation, the corresponding
critical exponents can be easily extracted from the low-energy effective theory
of light modes in $S^1_{\beta} \times {\bf R}^d$. 
The strength of the 
infrared singularity is characterized by the ``effective dimension''
$d_{\rm eff}$, the number of spatial dimensions that the centre of mass
of the string propagating in the loop is able to probe. 
Morally speaking, for a standard (open or closed)
propagating string state with $m^2_{\rm eff} \sim \beta-\beta_c$ we
have at one loop
\eqn\estione{
F(\beta)_{\rm 1-loop} \sim \int_0^{\infty} {ds\over s} \int d^{d_{\rm eff}}
{\vec p} \,\,
e^{-s({\vec p}^2 + \beta-\beta_c)} \sim (\beta-\beta_c)^{d_{\rm eff}/2}. }
Actually, this case is special because of the extra $1/s$ in the
moduli measure. At higher order ($2g+N \geq 2$)
we must estimate infrared divergences in
propagators rather than in the logarithm of a determinant\foot{This
can be seen alternatively as due to the fact that
there are no residual Killing vectors associated
with higher-order Riemann surfaces.}.
From the corner of the moduli space with $n$ soft loops of the same
type, $n\leq g$ or $N-1$, we get
a contribution
\eqn\tiotro{
F(\beta)_n \sim \lambda^{2g-2} \,
\left[ \int_0^{\infty} ds \int d^{d_{\rm eff}} {\vec p} \,\,
e^{-s ({\vec p}^2 +\beta-\beta_c )} \right]^n \sim
\lambda^{2g-2} \,  (\beta-\beta_c)^{n(d_{\rm eff}/2 -1)}.}

As we saw,
in the large-$E$ limit the leading term comes from 
the smallest critical exponent. From \tiotro\ we see that we will
run into trouble whenever $d_{\rm eff}\leq 2$, because then
an instability takes over in the sense that the singularity gets harder
as we increase the order of the diagram. This would indicate that the 
weak-coupling 
expansion breaks down at high energies, since higher-order Riemann
surfaces would dominate the large energy limit. As we will see later,  
this is exactly what happens in the one-instanton sector of the bosonic
string, where $d_{\rm eff}=0$ for ``soft strips'' 
(the open strings endpoints are stuck at the instanton
location). More generally, if we are
doing thermodynamics in the background of a fixed $p$-brane, we will find
problems for $p=0,1,2$.

On the other hand, for closed string soft loops,  
$d_{\rm eff}=d>2$. In the perturbative (zero-instanton) sector,  
these are the only types of handles present and therefore,  
for higher genus surfaces, the leading term in $\Omega(E)$ comes
from a Riemann surface with a 
single ``soft handle''. Incidentally, the effective
dimension for the interaction contributions ($g>1$) is two units smaller
than the one for $g=1$, so
these diagrams could dominate the torus contribution if the energy
is high enough, in spite of their suppression by powers of the (small) string
coupling.

As  is usual in instanton calculus, all physical quantities have to be 
expressed in a weak-coupling expansion around each instanton sector. This
entails the following form for the partition function
$${\cal Z}(\beta)=\sum_{n=0}^{\infty}\lambda^{2n}{\cal Z}_{2n}(\beta)+
e^{-|Q|/\lambda}
\sum_{m=0}^{\infty}\lambda^{m}
{\cal Z}_{m}^{'}(\beta)+{\cal O}(e^{-2|Q|/\lambda} ).  
$$
Using the linear relation \lapinv\ between the partition function and
the density of states we obtain the representation for 
$\Omega(E)$ in the dilute-gas approximation: 
\eqn\den{
\Omega(E)=\sum_{n=0}^{\infty}\lambda^{2n}\Omega_{n}(E)+
e^{-|Q|/\lambda}\sum_{m=0}^{\infty}\lambda^{m}\Omega_{m}^{'}(E)
+{\cal O}(e^{-2|Q|/\lambda}) . 
}
Then, the contributions of the one-D-instanton sector to $\Omega(E)$ in 
eq. \den\ are defined by
$$
\Omega_{m}^{'}(E)=\int_{c-i\infty}^{c+i\infty}{d\beta\over 2\pi i}
\,{\cal Z}_{m}^{'}(\beta)\,e^{\beta E}=
\int_{c-i\infty}^{c+i\infty}{d\beta\over 2\pi i}
\left[e^{-\beta F(\beta)_{\rm pert}-
\Gamma'_{1}(\beta)}\right]_{m}e^{\beta E}
,$$
where the subscript $m$ indicates that we are only retaining the coefficient
multiplying $\lambda^{m}e^{-|Q|/\lambda}$ in the expansion in powers of 
$\lambda 
$, and
$$
\eqalign{\Gamma'_{1}(\beta)=
\sum_{N=1}^{\infty}\sum_{h=0}^{\infty}{\lambda^{2h+N-2}\over N!}
W_{g,N}(\beta)
}
- {W_{0,1} \over \lambda}. $$
Since neither $F(\beta)_{\rm torus}\equiv F_{1}(\beta)$ nor 
$W_{0,2}(\beta)$ are weighted
by $\lambda$, we should keep them in the full exponential,
 multiplying all 
terms
in the weak-coupling expansion; the generic term 
${\cal Z}^{'}_{m}(\beta)$ will then be 
of the form
\eqn\genq{
{\cal Z}_{m}(\beta)=
e^{-\beta F(\beta)_{\rm torus}-{1\over 2}W_{0,2}(\beta)}\left[
\sum_{r,s}D_{(g_{i},h_{i},N_{i})}
\,F_{g_1}(\beta)\ldots F_{g_r}(\beta)
W_{h_1,N_{1}}(\beta)\ldots W_{h_{s},N_{s}}(\beta)\right]
,}
where $D_{(g_{i},h_{i},N_{i})}$ are some combinatorial factors and the sum
is restricted to $g_{i}>1$, $2h_{i}+N_{i}>2$, and
$
\sum_{k=1}^{r}(2g_{k}-2)+\sum_{\ell=1}^{s}(2h_{{\ell}}+N_{{\ell}}-2)=m$.   
In general we find that the contributions to the energy density in
the one-instanton sector mix Riemann surfaces with and without boundaries. 
This has to be so, since both connected and non-connected
Feynman--Polyakov diagrams appear in the weak-coupling expansion of 
${\cal Z}(\beta)=\exp[-\beta F(\beta)]$.

Using the techniques of Appendix B, we can get an estimate  
of the coefficients $\Omega_{m}^{'}(E)$
in the expansion of the density of states. The first term in the 
one-instanton sector is
\eqn\om{
\Omega_{0}^{'}(E)=\int_{c-i\infty}^{c+i\infty}{d\beta\over 2\pi i} 
\,e^{-\beta F(\beta)_{\rm torus}-{1\over 2}W_{0,2}(\beta)}\,
e^{\beta E}
.}
The first singular point contributing to this integral comes from
the leading singularity of $F(\beta)_{\rm torus}$, the Hagedorn inverse
temperature $\beta_{H}$. As a matter of fact, we know that $W_{0,2}(\beta)$
is regular at
$\beta_{H}$ (its rightmost singularity in the complex $\beta$ plane
lies at $\beta_{H}/2$), so,
modulo overall numerical factors, we get a term analogous to the leading
contribution of the perturbative part of $\Omega(E)$, except for the  
exponetial suppression by the  instanton action. Putting the
perturbative and non-perturbative terms together,
we end up with the following leading behaviour for $\Omega(E)$ as 
$E\rightarrow\infty$ (up to the one-instanton level)  
\eqn\leaom{
\Omega(E)\sim \left(C_{0}+C_{0}^{'}e^{-|Q|/\lambda}\right)
\,{e^{\beta_{H}E}\over
E^{{d\over 2}+1}},  
}
where $C_{0}$ and $C^{'}_{0}$ are two numerical constants.
The conclusion is that the only effects of D-instantons in the leading 
large-$E$ behaviour of $\Omega(E)$ is in changing the overall $E$-independent 
normalization\foot{
Actually, the fact that there is no D-instanton-induced singularities
at $\beta_{H}$ implies that the structure of the correction would be
the same for all the coefficients $C_{g}$ in a perturbative expansion
of $\Omega(E)$.}.
This only amounts to a constant shift in the microcanonical
entropy and thus has no consequences for the thermodynamics.
More interesting for the physics is the contribution to the
inverse Laplace transform coming from the first singular points 
of ${\cal Z}(\beta)$
to the left of the Hagedorn point,  $\beta_{c}=\beta_{H}/2$,
which is common to both $F(\beta)_{\rm torus}$ and $W_{0,2}(\beta)$.

Here it is convenient to split the discussion between the purely
bosonic case and the IIB D-instantons that will     be considered in
the next subsection. 
Since infrared
divergences in the Dirichlet sector have a vanishing effective
dimension, the one-loop open
 string diagram leads to a logarithmic singularity,  
as we already saw in eq. \dive. 
 When
defined with the natural normalization inherited from T-duality, the
full exponential  of the cylinder diagram 
 entering \om\ has a simple critical
exponent; we
simply find
\eqn\rat{
e^{-{1\over 2} W_{0,2} (\beta)} \sim (\beta-\beta_c)^{-1/2} . 
}
For $\exp[-\beta F(\beta)_{\rm torus}]$ we give the general expansion 
in eq. (B.1).

To complete the analysis we need to study the polynomial in \genq.
The critical behaviour of $F_{g}(\beta)$ is given in eq. \tiotro\ and
for $W_{h,N}(\beta)$ we have that, above the annulus diagram, all
infrared singularities
in the Dirichlet open string channel are governed 
by singular propagators. Now, however, the effective dimensions
of the infrared singularity is $d_{\rm eff}=0$, so
for $n$ soft strips ($n< N$) we have
\eqn\sfstrips{
W_{g, N} (\beta) \sim \left[ \int_{0}^{\infty} dt \, e^{-t(\beta-\beta_c)} 
\right]^n \sim (\beta-\beta_c)^{-n}.}
Because of the rational critical exponent in \rat\ we have that,
contrary to the perturbative case, now it is the single-valued part in
 (B.1) the
one contributing to the integral along the branch cuts.  
From (B.1), \rat\ and \sfstrips\ we find that ($m=2h+N-2$): 
$$
{\cal Z}^{'}_{m}(\beta)=
e^{-\beta F(\beta)_{\rm torus}-{1\over 2}W_{0,2}(\beta)} W_{h,N}+
{\rm less\,\,divergent\,\,terms}\sim (\beta-\beta_{c})^{-n+{1\over 2}}, 
$$
where $n<N$ is the number of soft strips; as a consequence the leading 
terms in the large-$E$ expansion in 
the instanton-induced density of states associated with the
singularity at $\beta_{c}=\beta_{H}/2$ will be of order
\eqn\finsts{
\lambda^{2g+N-2}\, e^{-|Q|/\lambda} \, E^{n-{1\over 2}} \, 
e^{\beta_{H}E/2 }.  
}
A negative critical exponent with no bound! This means that, even being 
strongly suppressed by the string coupling constant, the
contribution from surfaces with an arbitrarily large number
of boundaries will be the ones
dominating asymptotically the high-energy regime at the first subleading
singularity, $\beta_c = \beta_H /2$; moreover, since the critical exponent
is negative they
tend to make the specific heat positive. 
It is, however, difficult to decide 
{\it a priori} to what extent
\finsts\ is quantitatively important, because these terms  are down 
with respect to the contributions from the Hagedorn point \leaom\ by
two exponential suppression factors: one in coupling $e^{-|Q|/\lambda}$ and,
the most important, in energy $e^{-\beta_H E/2}$.

It is perhaps more  natural 
to interpret \finsts\ as a pathology of the bosonic string D-instanton
expansion. The fact that $n$ has no bound means that the large-$E$ limit
does not commute with the weak-coupling expansion and therefore we must
regard \finsts\ as a severe ``infrared catastrophe" of the dilute D-instanton
gas of the bosonic microcanonical ensemble. Notice that this pathology
is not a general problem of string perturbation theory, but rather 
a specific issue of the Dirichlet-instanton construction, since it is
only the Dirichlet channel singularities that produce negative unbounded
critical exponents.  

We can summarize the situation by saying that whenever a given Riemann
surface degeneration produces a positive specific heat critical exponent,
then the weak-coupling expansion and the high-energy limit fail to  
commute, rendering the whole dilute instanton approach useless. 
In other words, the occurrence of a negative specific heat phase at
high energies seems to be characteristic of the weak-coupling expansion,
even when supplemented by a dilute gas of instantons. In the next
paragraphs we try to argue that such pathologies are absent for
the type-II case, and therefore $dilute$ 
 D-instantons would not induce any
qualitative changes in the standard perturbative lore.  

\subsec{Type-IIB Strings}

For a number of reasons, 
we expect the type-II D-instanton gas to be  much better behaved.
First of all, the theory is (modulo rigour) finite at zero temperature.
In the bosonic case, the instanton contributions to the vacuum energy 
$\Lambda \equiv F(\beta =\infty)$ are notoriously singular:  
\eqn\vacen{
\Lambda = \Lambda_{\rm perturbative}
 - e^{-|Q|/\lambda} \,e^{-\Gamma'_1 (\beta=\infty)}
 + {\cal O} (e^{-2|Q|/\lambda}).}
We see that, due to the exponentiation effect of the instanton
calculus, the subtracted free energy 
 $F(\beta)-\Lambda$ is still afflicted from the 
zero-temperature tachyons running in handles contributing to 
$\Gamma_1' (\beta =
\infty)$. In the supersymmetric case, on the other hand,
both $\Lambda_{\rm perturbative}$
and $\Gamma'_1 (\infty)$ should vanish;  
 a space-time fermionic zero mode
develops in front of the full non-perturbative contribution, since 
 the D-instanton breaks half the supersymmetries of the type-II
vacuum (see Appendix A). 

Another property of the type-IIB strings  that makes the instanton
expansion much better behaved is the absence of  classical
interactions between instantons. In spite of this, instantons and 
anti-instantons do interact with a Coulomb tail at long distances
and, at the string scale, the interaction between them blows up in
a Hagedorn-like fashion. As we explained above, this 
problem is avoided 
by working with explicit hard cores of radius  
$\ell_{0}=\beta_{H}/2$. There is, in any case, a potential source of 
temperature-dependent divergences due to a winding open string joining
a D-instanton with an anti-D-instanton separated by a distance $\ell$. 
At finite temperature, in the imaginary time formalism,
 one can view such a 
configuration as a periodic array of instantons with separation $\beta$ 
interacting with a similar parallel array of anti-instantons. 
Temperature-dependent
singularities in their interaction potential appear whenever there is an
open string stretching between these two arrays with length smaller
than $\beta_{H}/2$. Because we are in flat Euclidean space, the length
of such a string will be simply $\ell^{2}+n^{2}\beta^{2}$, with $n$ an
integer, the winding number; because we are placing the
restriction $\ell>\ell_{0}=\beta_{H}/2$, the presence of
the hard core guarantees that there are no divergences for any value
of the inverse temperature $\beta$.

Let us recall the detailed structure of the cancellation in the
interaction between instantons.
The world-sheet Hamiltonian in the Dirichlet open string channel
takes the form $\Delta_{s}  = (L_0 -a)_s$ with $a_s $ the standard
intercept for the corresponding spin structure: $a_{NS} = 1/2$,  $
a_R = 0$, and $L_0^{(s)} = {n^2 \beta^2 \over 4\pi^2 \alpha'} + (
{\rm oscillators})^{(s)}$. The appropriate GSO projection leading
to a vanishing loop diagram is ($w=e^{-t}$)  
\eqn\bps{
\eqalign{ 
W_{0,2} =& -{1\over 4}\int_{0}^{\infty} {dt\over t}\Big[ {\rm Tr}_{NS} 
\left(1-(-1)^F \right) e^{-t\Delta_{NS}} - {\rm Tr}_{R} \left(1-(-1)^F
\right) e^{-t\Delta_R}    
\Big] \cr  
=&-{1\over 4} \int_{0}^{\infty} {dt\over t} \sum_{n\in {\bf Z}} e^{-t{n^2  
\beta^2 \over 4\pi^2 \alpha'}} \Bigg[ w^{-1/2} \prod_{m=1}^{\infty} 
\left( {1+w^{m- 1/2} \over 1-w^m }\right)^8  \cr & 
-w^{-1/2} \prod_{m=1}^{\infty} 
\left( {1-w^{m-1/2} \over 1-w^m}\right)^8 -16\prod_{m=1}^{\infty} \left(
{1+w^m \over 1-w^m }\right)^8 \Bigg]
=0.}}
Upon modular transformation $s=-\log q = -2\pi^2 /\log w = 
2\pi^2 /t$, we  obtain the appropriate GSO projection in the
closed string channel, where the vanishing of \bps\ is explicitly
seen as a result of the BPS balance between the purely bosonic
states coupling to the (bosonic) D-instanton. 

Alternatively, using a Green--Schwarz formulation with the standard
${\bf 8_s}$ of $SO(8)$ of space-time fermions $S^a (z,{\bar z})$, we
have a single open string vacuum with the structure $|{\bf 8_v}\ket 
\oplus |{\bf 8_c}\ket$. The first term represents bosonic ground
states and the second one fermionic ground states, i.e. 
$(-1)^F |{\bf 8_v}\ket = |{\bf 8_v}\ket$ and $(-1)^F |{\bf 8_c}\ket
=-|{\bf 8_c}\ket$, and $F$ now represents the space-time fermion number.
Then we have the representation
\eqn\ot
{W_{0,2} = -{1\over 2} \int_{0}^{\infty} {dt\over t} {\rm Tr} (-1)^F 
e^{-t\Delta_{GS}} = -{1\over 2}\int_{0}^{\infty} {dt\over t} 
\prod_{m=1}^{\infty} \left( {1+w^m \over 1+w^m }\right)^8 \cdot
{\rm Tr}_{\rm vacua} (-1)^F =0.}
Formally, we can envisage a Ward identity in the following way: let
${\cal H}_{\rm open}^D$ be the physical Hilbert space of the open
Dirichlet string, and split it into bosonic and fermionic components\foot{
The notion of space-time fermion in the zero-dimensional space-time
of the open string channel is provided by the continuation from higher
D-branes. In particular, at the massless level, we just have the dimensional
reduction of $D=9+1$ supersymmetry down to zero dimensions.}   
${\cal H}_{B} \oplus {\cal H}_F$, related by the unitary operator ${\cal 
U}_{ss}$ carrying 
the representation of supersymmetry in the physical Hilbert space 
${\cal U}_{ss} {\cal H}_B = {\cal H}_F$. Now
\eqn\wi
{{\rm STr}_{\cal H}\, e^{-t\Delta} \equiv {\rm Tr}_{{\cal H}_B}   
e^{-t\Delta} - {\rm Tr}_{{\cal H}_F} e^{-t\Delta} = {\rm Tr}_{{\cal H}_B}
\left( 
e^{-t\Delta} - {\cal U}_{ss}^{-1} e^{-t\Delta} {\cal U}_{ss} \right) =0,} 
since $[{\cal U}_{ss}, \Delta ] =0$.
Heuristically, all perturbative non-renormalization theorems in string
theory have this structure. If a Riemann surface with at least one
closed loop (in a closed or open string channel) has an unbroken
space-time supersymmetry, implemented by the operator ${\cal U}_{ss}$, then
the diagram vanishes by an analogous argument. One would just
open the loop inserting unity and write the full diagram in operator
form as
\eqn\wid
{{\rm STr}_{\CH^D_{\rm open}}
\, [\Delta^{-1} \, {\cal O}(\Sigma') \, \Delta^{-1} ]   
= {\rm Tr}_{{\cal H}_{B}} [ \Delta^{-1} \, ({\cal O}(\Sigma') -
{\cal U}_{ss}^{-1} {\cal O}(\Sigma') {\cal U}_{ss} ) \, \Delta^{-1} ], }
where ${\cal O}(\Sigma')$ represents the operator insertion of the
rest of the Riemann surface. This clearly vanishes if $[{\cal U}_{ss},
{\cal O}(\Sigma')] =0$, i.e.  if the Riemann surface has the
unbroken ${\cal U}_{ss}$-supersymmetry\foot{It is important that
we have at least one closed loop in the Riemann surface, so that for
example there is no vanishing theorem for the disk diagram. This is
just fine, since the disk determines the D-instanton action, and
it is certainly non-zero in the type-IIB theory.}.
In a covariant formalism we must realize ${\cal U}_{ss}$ in terms
of a contour integral of the corresponding supersymmetric current, and then
we would prove \wid\ by a contour deformation argument. To be more 
precise, taking into
account the spin structures, the right-hand side of \wid\ is not
always zero but rather leads to 
a boundary term on the perturbative moduli space
of Riemann surfaces, which we tacitly discard in these conceptual
digressions.

The cylinder or annulus diagram vanishes in \bps\ or \ot\ at
finite temperature, exactly as it does at zero temperature. In other 
words, this particular diagram is ``supersymmetric", even if we
know that supersymmetry is broken by the thermal boundary conditions. Notice
that, in individual diagrams, supersymmetry
is broken because boundary conditions
for fermionic loops
 now differ  from those of    bosonic loops (i.e. 
antiperiodic rather than periodic). However, in the cylinder diagram,
the only states  propagating around the thermal loop are the
closed string states (it is the cylinder  that wraps around
the thermal circle before it ends again at the instanton location). But
all states in the closed string channel are bosonic, because the
D-instanton 
 boundary state
\bs\ is bosonic. 
This  is right as long as the temperature is not too
small, in which case we have to consider the fermionic quasi-zero modes
discussed in Appendix A, inducing fermionic boundary states. However,
in the vicinity of the critical temperature, such modes have a gap
of the order of the string scale, and only the bosonic D-instantons
should be considered.  

Since the cylinder diagram is purely bosonic in the closed string
channel, the GSO projection is the same as in the zero-temperature case,
and we therefore have the standard vanishing result. Clearly, as soon as
 we have at least
one closed-string loop which could propagate fermions around the thermal
circle,  the diagram does break supersymmetry. However, the D-instanton is
still coupling only to bosonic states, so that if  we isolate the
open string strip between two boundaries we still can represent the
amplitude as in                                         
\wid
\eqn\widd
{W_{g,N} = {\rm STr}_{\CH^D_{\rm open}}
\, \left( \Delta^{-1} \, {\cal O} (\Sigma_{g, N-2}) \,
\Delta^{-1} \right), }
where now ${\cal O}(\Sigma_{g, N-2})$ represents the operator insertion
of the rest of the Riemann surface, having exactly one less open-string
loop, which accounts for two holes. Now, as long as we have 
 $g>0$, some closed-string fermions will wrap the
thermal circle and therefore $[{\cal U}_{ss}, {\cal O}(\Sigma_{g, N-2})]
\neq 0$, i.e. $W_{g,N}$ cannot be argued to vanish on the basis of
a supersymmetric Ward identity. However, by world-sheet locality, it is still
true that no fermions run in the closed channel of the selected strip.
So, locally, the GSO projection in this part of the Riemann surface
must be the same as at zero temperature. This means that the open
string tachyon in the NS sector triggered at $\beta_c = \beta_H /2 =
\pi \sqrt{2\alpha'}$ by $\Delta^D_{\rm open} (NS)
 = {\beta^2 \over 4\pi^2 
\alpha'} -{1\over 2} =0$, is again projected out  by the same 
GSO projection applied in \bps. Working in the 
Green--Schwarz formalism,
the propagators entering \widd\ would have the structure
$
\Delta_{GS} = {n^2 \beta^2 \over 4\pi^2 \alpha'}  
-a + {\rm osc}$,  
with  integrally moded fermions $S^a$ and then $a=0$.

So the final conclusion is that in the type-IIB superstring all 
temperature-dependent divergences in the one-instanton sector
are associated exclusively with the degeneration of closed string handles.
This means that there are no new string states becoming massless
at the Hagedorn temperature besides the tachyonic winding
mode already present in the zero-instanton sector. The obvious consequence
is that we get the same critical exponents for ${\cal Z}(\beta)$
as in the zero-instanton sector;  the only change
in the large-$E$ expansion of $\Omega(E)$ then 
is a modification of  the coefficients
of the asymptotic series, much in the fashion of eq. \leaom.
Because of the absence of singularities in the partition function
associated with D-instantons, there are no terms in $\Omega(E)$ contributing
to positive values of the specific heat. This makes the weak-coupling 
expansion well defined for the type-IIB strings, but on the other hand
seems to indicate that the introduction of D-instanton effects
does not imply any qualitative modification in the thermodynamics of
type-IIB superstrings near the Hagedorn temperature, as it is given by
string perturbation theory in the microcanonical ensemble.

\newsec{Mean Field Analysis of the Hagedorn Transition}

Having studied the effects of D-instantons in the microcanonical 
description of string thermodynamics, we now turn to an analysis
of the instanton effects on the Hagedorn transition.
In the previous section we saw  that, at least for the case
of the type-IIB superstring, single 
D-instanton effects do not introduce
new divergences; therefore, we can  assume the winding
closed string tachyon to be the only  relevant order parameter of the
Hagedorn phase transition.
In what follows, we will include dilute instanton effects in 
the mean field approximation of \refs\raw,
which leads to a first-order transition in perturbation theory. 

\subsec{Mean Field Approximation}

Following Atick and Witten \refs\raw,
 we consider the effective long-wavelength
dynamics  of the complex tachyon field
denoted by $\sigma$,
 which is assumed to nucleate the phase transition. For 
this purpose we use a mean field approach, considering $\sigma$ as
a smooth background field in the spatial dimensions, and we
integrate out fluctuations for all the approximately massless degrees
of freedom. At large spatial scales, only non-derivative interactions are
important and we have the following effective nine-dimensional 
Lagrangian as a starting
point
\eqn\eee{
\eqalign{\CL_9 =& -\sigma^{*}
 {\vec \partial}^2 \sigma + m_{\rm eff} (T)^2 \sigma  
\sigma^{*} +\lambda^2 T \alpha'^3 c\, (\sigma \sigma^*)^2 \cr &
 - {1\over 2} \sum_f \varphi_f (-{\vec \partial}^2
 + m_f^2 ) \varphi_f +
\lambda \alpha'  \sqrt{T} \sum_f c_f \sigma^{*} \sigma \varphi_f + ...
}}
where we have included the quartic self-coupling of the tachyon
induced at tree-level, $c \sim \bra V_{\sigma} V_{\sigma} V_{\sigma^*}
V_{\sigma^*}\ket $ is of order unity and positive, since we do not
expect instabilities at zero temperature.  The
 ellipsis stands for higher-dimension couplings and 
higher-derivative
 terms. The effective mass squared  $m_{\rm eff}^2$ vanishes
linearly at the critical temperature\foot{In the type-IIB theory we have
 $m_{\rm eff}^2 = {\beta^2 \over
4\pi^2 \alpha'^2} - {4\over \alpha'}$ at tree-level.}, and  the sum in 
\eee\   runs over the set
of almost massless fields at the singularity (radiatively or 
instanton-induced masses).  
Integrating out the fields $\varphi_f$ at tree-level we obtain an
effective potential\foot{We  
 can tune the string coupling $\lambda$ to be artificially
small. In fact, we must do so in order to avoid the Jeans instability,
which would invalidate any thermodynamical approach \refs\raw.}          
$$
V_{\rm eff} (\sigma, \sigma^*)
 = m_{\rm eff} (T)^2 \sigma^{*} \sigma +\lambda^2 T \alpha'^3 c 
(\sigma \sigma^*)^2 - {1\over 2}     
\lambda^2 T \alpha'^2  \sum_f c_f^2
 \bra \sigma^* \sigma |(-{\vec \partial}^2 +m_f^2 )^{-1}
| \sigma^* \sigma \ket
$$
and in the infrared limit we are left with the contact term  
\eqn\cont{
V_{\rm eff} (\sigma, \sigma^*) \rightarrow 
 m_{\rm eff} (T)^2 \sigma^* \sigma + \lambda^2 T \alpha'^3 c 
(\sigma \sigma^*)^2  - {1\over 2} 
\lambda^2 T \alpha'^2  \sum_f {c_f^2 \over m_f^2 } (\sigma^* \sigma )^2 . 
}

The character of the phase transition is therefore determined by the
relative values of the ratios of tree-level couplings to masses
$c_f^2 / m_f^2$. Fields $\varphi_f$ with a real coupling $c_f$ to the
NS--NS tachyon $\sigma$ contribute  negative quartic terms  
to \cont, and if dominating it  follows that  
the  phase transition is first order, since an instability
to the condensation of the field $\sigma$ sets in when we still have
$m_{\rm eff}^{2} (T) >0$. Higher-order terms are expected to stabilize
       $V_{\rm eff}$ at large field strength. 
If one of the fields $\varphi_f$ is exactly massless, then $m_f^2$ should
be replaced by some infrared cutoff, say the inverse volume of the
spatial box. This is perfectly reasonable since we are simply exhibiting
the qualitative character of the transition as first order as opposed
for example to a second-order transition, where the order parameter
only condenses when it becomes tachyonic.

On the other hand, if $c_f$ is imaginary then $\varphi_f$ is a sort
of ``scalar gauge field" with respect to which $\sigma$ has charge
$|c_f|$. Such terms induce a positive quartic term in \cont\ and,  
if dominating, we have a second-order transition instead.
Notice that, since we are assuming $c>0$, the bare potential tends to
produce a second-order scenario.

The analysis of \refs\raw\ dealt only with the dilaton. In fact, it
is easy to argue that the other universal NS--NS massless field,
the antisymmetric tensor $B_{\mu\nu}$, is irrelevant for the
infrared limit in \cont. The fields $\sigma$ (resp. $\sigma^*$) have
winding number $+1$ (resp. $-1$) so that they have $\pm$ charge with
respect to the nine-dimensional gauge field $A_i = B_{i0}$. However,
the corresponding electromagnetic current $\sigma^* {\vec \pt} \sigma$
contains a derivative, which makes it irrelevant in the infrared. 

These considerations lead us to focus just on the axion--dilaton scalar
sector of the type-IIB theory. In perturbation theory the dilaton 
couples at tree-level to the tachyon field, $c_{\phi}$ is of order unity, 
and a radiative mass is generated at one-loop order $m^2_{\phi} \sim
\lambda^2 /\alpha'$, for temperatures close to the critical Hagedorn        
point\foot{For $T \leq \alpha'^{-1/2}$ we can easily estimate the self-energy
correction in the low-energy effective theory \eee\ keeping also the
dilaton Kaluza--Klein modes of mass $\sim T$. A simple dimensional analysis
of the loop diagram gives $m^2_{\phi} \sim \lambda^2 T^6 \alpha'^{2}$.
More detailed treatments of mass corrections in the full-fledged 
string theory can be found for example in \refs\rminahan.}.
Then the dilaton contribution to quartic terms in \cont\ is
\eqn\dpot
{\delta V_{\rm dilaton} \sim - (\alpha')^{5/2} (\sigma \sigma^*)^2 }
for $T\sim \alpha'^{-1/2}$, and we see that this term, at weak coupling,   
 clearly dominates the bare tachyon potential and settles the
scenario of a first-order transition for the case of the bosonic or
heterotic string. Since both $c_{\phi}$ and $m^2_{\phi}$ are
sizeable in perturbation theory, instanton corrections make a negligible
contribution to \dpot.

In the type-IIB theory this is not the end of the story because
of the axion field.
Here both the coupling to the tachyon $c_a$ and the induced mass
$m_a$ vanish in perturbation theory, as a result of the classical
$SL(2,{\bf R})$ symmetry 
of the ten-dimensional type-IIB theory, which in particular
involves $a\rightarrow a+ {\rm constant}$, and forbids non-derivative
couplings of the axion in perturbation theory. This symmetry is broken
by D-instantons and in general by the existence of D-branes to a
discrete $SL(2,{\bf Z})$ symmetry, so that  periodic couplings of the
axion can be generated at the non-perturbative level. Since the 
derivative of the axion $\pt_{\mu} a$ behaves like a $U(1)$ field strength
--in particular it is dual to the 9-form field strength $F_9$
coupling to 7-branes-- it is at least possible that a non-perturbatively
generated tachyon--axion coupling would have $c_a$ pure imaginary.
 
We see that the sign and magnitude of the  ratio $c_a^2 / m_a^2$
is completely determined by non-perturbative physics, and  we conclude
that the D-instanton analysis is crucial to decide the order of the
transition in the type-IIB theory.

\subsec{Axion Couplings}
   
The classical symmetry $a\rightarrow a+ {\rm constant}$, is embodied
in type-IIB 
string perturbation theory in the fact that R--R vertex operators
are ``longitudinal" or, more precisely, correspond really to field
strengths rather than to the ``photon" fields, and correlation
functions vanish at zero momentum\foot{ For example, the axion
vertex operators in the $(q_1, q_2)$ picture read
$$
V_a^{(q_1,q_2)} = p_{\mu} (\gamma^{\mu})_{AB} V^A_{q_1} (z) V^B_{q_2}
({\bar z})
$$
and, in the canonical $q=-1/2$ picture,  
$
V^A_{-1/2} (z) = e^{-\phi(z)/2} S^A (z) e^{ipX(z)}  
$, 
where $\phi(z)$ here denotes the bosonized ghost field and has no 
relation with the  dilaton. Then, a simple OPE check shows that
mass corrections $ \bra V_{RR}  V_{RR}  \ket $   
 and vertex amplitudes $\bra V_{RR} V_{\sigma} V_{\sigma^*} \ket$ 
induce derivative interactions after we integrate the vertex operators
over the world-sheet,  so that  all amplitudes vanish at zero
 momentum.}.
Even more generally, this follows from separate left--right conservation
of the world-sheet fermion number.

Such arguments could in principle 
fail when  considering D-instantons, since  
 the integral over instanton locations might pick up surface terms
coming from the derivative couplings, leading to a constant in the
zero-momentum limit. Now, there is no  {\it a priori} reason  
for amplitudes like
\eqn\tress{
J_{\rm coll} \,A \, 
 \, e^{-|Q|/\lambda}\, \lambda^2 \, \int dx_0 \,\bra V_{\sigma}
 V_{\sigma^*} V_a |I, x_0 \ket_{\rm disk} 
}
contributing to cubic $\sigma \sigma^* a$ couplings to vanish in the
zero-momentum limit.
In fact, the leading instanton contribution to the cubic axion--tachyon
coupling has no extra powers of the string coupling 
constant and it is given by
\eqn\lead{  
J_{\rm coll} \,  
A \,e^{-|Q|/\lambda} \int dx_0 \,\bra V_{\sigma}|I,x_0 \ket_{\rm disk}
\,\bra V_{\sigma^*} | I,x_0 \ket_{\rm disk} \,
\bra V_a | I, x_0 \ket_{\rm disk}, }
which clearly  vanishes: the net winding
number introduced in the disk by the tachyon vertex operators cannot
flow through the instanton, since we have Dirichlet 
boundary conditions on the boundary of the disk for {\it all} directions.

 The next correction scales with an overall power of $\lambda$ and has
the form
\eqn\next{
J_{\rm coll}\, A\,  e^{-|Q|/\lambda}\,\, 
 \lambda \int dx_0 \,\bra V_{\sigma} V_{\sigma^*}
|I, x_0 \ket_{\rm disk} \,\bra V_a |I, x_0 \ket_{\rm disk}. } 
In this case, it is clear that we have a non-trivial result because the
$\sigma$--$\sigma^*$ pair  couples dominantly
 to the D-instanton via dilaton
exchange, and the axion couples directly to the D-instanton.

 On general
grounds, the disk 
``tadpole" or          one-point function of a vertex operator in
the background of the instanton is related, after propagator amputation,
to the classical profile of the associated 
 field in the instanton configuration. For example, for a scalar field
$\Phi (x)$  we can write,   
 in a somewhat symbolic fashion,   
\eqn\sym{
(-\pt_x^2+m^{2}_{\Phi} )\Phi (x-x_0)_{c\ell} 
 \sim \int dp\, e^{ipx}\, \bra V_{\Phi} (p) |I, x_0
 \ket_{\rm disk}.  }  
This is interesting because the classical profile of the axion field
$a_{c\ell}$, which is purely imaginary,  enters \next\ directly, 
so that  we have a specific source for imaginary contributions to
the axion coupling $c_a$. 

A detailed analysis of the size and sign of the instanton-induced
axion--tachyon couplings is complicated. Fortunately for us, the
structure of \cont\ implies that, as long as an axion mass is
generated in the one-instanton sector, the dilaton dominance of
the phase-transition dynamics is ensured, and therefore the standard
first-order picture would not be modified. This follows from the fact
that, up to powers of the string coupling, 
 $|c_a |^2 \leq e^{-2|Q|/\lambda}$, because $c_a$ is at most generated
in the one-instanton sector. On the other hand, if the mass squared
is generated at the one-instanton level we have $m_a^2 \sim e^{-|Q|/
\lambda}$ 
and thus the relevant ratio
$$
{|c_a|^2 \over m_a^2} \leq e^{-|Q|/\lambda} . 
$$ 
This implies that the dilaton dominates independently of the sign of
$c_a^2$. The axion would be competitive only if its mass
were generated at the two-instanton level. In such a case the size of
the ratio $|c_a |^2 /m_a^2 $ would depend on the power dependence in 
the string coupling.  

     Now, from the general
discussion of section 2.3  we know that long-range Coulomb
 interactions of instantons without fermion zero modes  
 generate
periodic potentials for the axion. The discussion in Section 2.3 is
obscured by the fact that D-instanton gases do not really form a
plasma, because dilaton exchange exactly cancels the instanton-instanton
repulsion. The result was a runaway potential for the dilaton that   
is only stabilized at vanishing coupling. However, at finite temperature
this picture is only valid in the intermediate scales between the 
string scale and the induced mass of the dilaton $m^2_{\rm dil} \sim
\lambda^2 T^6 \alpha'^2 \sim \lambda^2 /\alpha' $, which can be
considerably lower than the string scale for weak coupling, even at
the critical temperature. Below this energy scale the dilaton exchange
is screened and we have a real Coulomb plasma. The runaway of the 
dilaton is in turn stabilized by the radiatively induced potential and
we end up with an axion mass at the one-instanton order      
\eqn\axm
{m_a^2 = 2\, A\, J_{\rm coll} \,Q^2 \, e^{-|Q|/\lambda} + ...}
so we conclude that the standard first-order scenario is not modified.

\newsec{Concluding remarks}

In the present paper
 we have tried to clarify the relevance of D-instanton effects
in string thermodynamics, with special interest on their influence in
the critical behaviour of the string collective. 
We have seen that D-instantons can induce strong infrared singularities
in thermodynamical quantities. Indeed, for purely bosonic D-instantons,
the whole instanton expansion seems ill-defined in the microcanonical
approach. On the other hand, type-IIB D-instantons are smooth within
the dilute approximation, although they do not  qualitatively modify  
the perturbative picture near the Hagedorn point. 

Still, the physics of dense ``liquids" containing
instantons and anti-instantons remains obscure, and the existence
of thermal singularities in the type-IIB string analogous to those found 
in the bosonic
case cannot be completely excluded. Anyhow, the corresponding critical
points are necessarily beyond the Hagedorn temperature, and it is
quite unlikely that they are physically relevant at a quantitative
level. This would be very odd on physical grounds, because we
expect instanton physics to produce very small effects at weak
coupling, when competing with perturbative contributions. The opposite
situation would rather 
 be interpreted as a breakdown of the semiclassical
expansion.   

A partial exception is the contribution of the type-IIB axion
to the nucleation of the Hagedorn transition, since both its
mass and non-derivative couplings are generated by 
finite-temperature
 instantons. We have found in the mean field approximation
that the dilaton is still dominating and the first-order scenario
of \refs\raw\ is not modified at this level.    

It is somewhat disappointing that D-instantons do not seem to
solve the ``Hagedorn problem". However, in retrospect this is  
a rather natural conclusion in view of the above comments. It
means  that the change of degrees of freedom
taking place is more radical than a simple addition of 
 non-perturbative
semiclassical states.

Finite temperature remains an interesting arena for the study
of D-instantons and D-branes.  Specially interesting are the
 infrared effects that      
are otherwise forbidden by supersymmetry, such as    
 the scalar potentials
generated by long-range interactions.   
At finite temperature, supersymmetry is only broken ``softly",
because the gravitinos  remain in the spectrum with masses
of order $\beta^{-2}$. We have seen that this is important
for the consistency of the D-instanton expansion.

\newsec{Acknowledgements} 
It is a pleasure to thank E. \'Alvarez, L. \'Alvarez-Gaum\'e,  
 D.J. Gross, R. Emparan, I.R. Klebanov, M. Laucelli, T. Ort\'{\i}n,
 M.A.R. Osorio,
J. Puente-Pe\~nalba and F. Quevedo
 for useful discussions and suggestions. 
The work of  
 M.A.V.-M. was partially supported  by the 
Spanish Science Ministry and by a Basque Government
Postdoctoral Fellowship. 

\appendix{A}{Fermionic Quasi-zero Modes}
Fermionic collective coordinates are easily implemented in the
light-cone formalism where explicit expressions for the supersymmetry
charges are available \refs\rmipaper\refs\rgg\refs\rhama. The full
super-boundary state is given by
\eqn\bsf{ |I,x,\theta\ket = e^{i\theta Q_{\rm broken}} \,|I,x\ket,}
where $|I,x\ket$ is the bosonic boundary state \bs, and the fermionic
superpartners are generated by the broken supersymmetry in the standard
fashion. For a given type-IIB instanton there are 16 independent
fermionic collective coordinates and, expanding \bsf, we obtain the
full multiplet of BPS-saturated instanton boundary states, entirely
analogous to an instanton superfield in the treatment of \refs\ritep.
The unbroken supersymmetry fixes the measure to be  
\eqn\mea{ d\mu = d^{10} x\, d^{16} \theta \, J_{\rm coll}.}
If $\theta$ represent quasi-zero modes, we induce the corresponding
fermion exchange interactions upon $\theta$-integration. For example,
in the I--A sector, the leading (cylinder) static interaction in
super-space is given by $\Gamma (x^+ - x^- ; \theta^+ , \theta^- ) 
= \bra x^- , \theta^- | (\Delta_{\rm closed} )^{-1} | x^+ , \theta^+
\ket$. Using \bsf\ and the supersymmetry algebra $\{Q_+ , Q_- \} = 
-i\sqrt{2} \gamma\cdot \pt$ we find
\eqn\red{ \Gamma (x^+ -x^- ;\theta^+ , \theta^- ) = 
e^{i\sqrt{2} \theta^- \gamma^{\mu} \theta^+ \pt_{\mu}} \, 
\Gamma_{I-A} (x^+ - x^- ), }
where the classical I--A overlap is given by (see for example        
\refs\rgreenf)  
\eqn\overl{ \Gamma_{I-A} (x-y) = -(2\pi)^4 \int_{0}^{\infty} 
{ds\over s^5}\, e^{(x-y)^2 \over 2\alpha' s} \, \prod_{m=1}^{\infty}
\left( {1+q^{2m} \over 1-q^{2m}}\right)^8 }
in the closed string channel, (here $q=e^{-s}$).
 Saturation of fermionic coordinates can
be done in several ways, depending on the number of disconnected 
world-sheets used. For example, with 16 cylinders connecting the instanton
and anti-instanton we have a product of 16 fermionic propagators with
the states $Q_-^{\alpha} |I_+ ,x^+ \ket$, $\alpha= 1,...,8,{\dot 1},...,
{\dot 8}$ running through each cylinder. With just one cylinder we
have the top (bosonic) state $Q_-^1 \cdots Q_-^8 Q_-^{\dot 1} \cdots
Q_-^{\dot 8} |I_+ , x^+ \ket$ running.

At finite temperature, antiperiodic boundary conditions on space-time
fermions lift all the supersymmetric zero modes, and therefore we
should not consider fermionic collective coordinates in the generic
situation. However, as $\beta\rightarrow\infty$ in the low-temperature
limit we should recover the zero-temperature results and in particular
the restoration of normalizable zero modes ensuring for example vanishing
vacuum energy. The corresponding zero modes are difficult to see by
a direct study of the individual diagrams $W_{1,N}$ in \unme. The reason
is that the Dirichlet construction only propagates free strings in
flat space, and the zero-mode singularity in the effective action
would appear only after a resummation of boundaries has been
performed, in the sense of eq. \resum. In any case, we can
derive the leading suppression factor by defining collective coordinates
for the quasi-zero modes at low temperature, in the sense of a
constrained instanton expansion.

The idea is simple: the family \bsf\ does not define classical solutions
at finite temperature, but it certainly defines approximate solutions
for large $\beta$. Therefore, we can still integrate over the whole
fermionic family \bsf, provided we include the classical interaction
with the thermal boundary conditions. As in the I--A sector, such
interactions are given by propagators of the states in the BPS multiplet.
In this case, a D-instanton at finite temperature is equivalent to an
array of D-instantons at distance $\beta$, and we must compute the
classical interaction between them induced by fermion exchange.

Denote the corresponding BPS boundary state by $(Q_-)^{\{p\}} |I_+ \ket$,
where $\{p\}$ denotes a subset of $p$ indices in the ${\bf 8_s} \oplus
{\bf 8_c}$ of $SO(8)$. Then, by the properties of the charges and the
explicit form of \bs\ we have $(Q_- |I_+ \ket)^{\dagger} = \bra I_- |
Q_+$. Therefore, a propagator wrapped around the thermal loop has the
form
\eqn\propt
{\left( (Q_-)^{\{p\}} |I_+ \ket \right)^{\dagger}
 \Delta_{\rm closed}^{-1}  
(Q_-)^{\{p\}} |I_+ \ket_{p} = \bra I_- |
(Q_+)^{\{p\}} \Delta_{\rm closed}^{-1} (Q_-)^{\{p\}} |I_+ \ket_p ,}
and indeed looks like an I--A interaction. The subindex means that 
fermions should be antiperiodic in the case that  
 $p$ is odd. Using the supersymmetry
algebra this is equal to   
\eqn\dett{  
 2^{p/2}\, \,{\rm det}_{\{p\}} 
(\gamma\cdot \pt) \,\, \Gamma_p (\beta) .} 
The determinant is over the subset $\{ p\}$ of Dirac indices, and
$
\Gamma_p (\beta) $ is the I--A overlap in the infinite array of
instantons. For $p$ even, it is just the periodic sum of the
elementary I--A overlaps in \overl. For $p$ odd we have
antiperiodic boundary
conditions, which is easily accomplished by introducing a
phase for odd wrappings of the form
\eqn\overf
{\Gamma_{p={\rm odd}} (\beta) = \sum_{n=0}^{\infty}(-1)^n \,
 \Gamma_{I-A} (n \beta) .}  
The function $\Gamma_{I-A}$ was written above \overl, and the 
antiperiodicity phase produces half integer modding after
Poisson resummation in $n$. 

Now, the determinant factor gives a term $(\pt_{\beta} )^p \, \Gamma_p$
for each cylinder. With $n_c$ cylinders such that $\sum_{i=1}^{n_c} p_i 
= 16$, and given the low-temperature scaling of $\Gamma_p$ we find 
that one-instanton effects are suppressed in the low-temperature limit
as 
\eqn\supr
{\prod_i \left( {\pt\over \pt \beta}\right)^{p_i} \, \Gamma_{p_i}
 (\beta) 
\sim \prod_i \beta^{-8-p_i} = \beta^{-8(n_c +2)} \leq \beta^{-24},}
 in agreement with our expectations.

\appendix{B}{Asymptotic form of $\Omega(E)$}

In this Appendix we will outline the calculation leading to the asymptotic
expansion of $\Omega(E)$ for large values of $E$ at one loop. Following 
ref. \ric\ we compute the inverse Laplace transform \lapinv\ by moving $c$ 
to the left and deforming the contour in order to avoid the singular points of
the integrand\foot{See \refs\rvet\ for a different treatment of the
large-$E$ limit.}. This demands
 the analytic continuation of ${\cal Z}_{0}(\beta)=
\exp[-\beta F(\beta)_{\rm torus}]$  to the whole complex $\beta$-plane.
This is a subtle step since it is known \refs\rov\ that the
 modular-invariant
representation of $F(\beta)_{\rm torus}$ is afflicted by singularities at
the self-dual length under $\beta$-duality, $\beta_{\rm{s-d}}$, 
 of the form
$|\beta-\beta_{\rm{s-d}}|^{a}$ with $a$ odd. Obviously, this kind of
singular behaviour does not allow an 
analytic continuation to complex values of
$\beta$. To avoid these
 problems we will work with the so-called S-representation
of the torus free energy, which is
 obtained by summing up the individual contributions
to the free energy from the fields contained in the string spectrum 
\refs\ranalogo and \alos. 
The expression so obtained can be analytically continued to complex values of
the inverse temperature and, as a bonus,
 there are no singular points except those
located on the real and imaginary axes.
 Actually, for real $\beta$ we have  
branch-point
 singularities
 located at $\beta_{n}=\beta_{H}/n$ ($n\in {\bf Z}$). Taking
all branch cuts to the left we decompose \lapinv\ into an integral along  
${\rm Re\,}\beta=c$ plus the integral along the contour $C$ surrounding the
branch cuts. The integral over the vertical contour will be of order
$\exp(cE)$, so we can write
$$
\Omega_{0}(E)=\int_{C}{d\beta\over 2\pi i}{\cal Z}_{0}(\beta)e^{\beta E}
+{\cal O}(e^{cE} ).  
$$
Let us assume first that $\beta_{H}/2<c<\beta_{H}$. In this case we can
expand ${\cal Z}(\beta)_{\rm torus}$ in a series of $(\beta-\beta_{H})$
 with  
a convergence radius equal to the distance to the closest singularity (i.e.
$\beta_{H}/2$). One can easily convince oneself that
\eqn\expaa{
{\cal Z}_{0}(\beta)\equiv e^{-\beta F(\beta)_{\rm torus}}=
\sum_{n=0}^{\infty} a_{n}^{(1)}
(\beta-\beta_{H})^{n+{d\over 2}}+\sum_{n=0}^{\infty} b^{(1)}_{n}
(\beta-\beta_{H})^
{n}
.}
The contour of integration $C$ is decomposed into $C_{\pm}$, parametrized as
$\beta=\beta_{H}+e^{\pm i\pi}$. From \expaa\ it is obvious that only the
multivalued part contributes to the integral, since the second term cancels 
out. After some straightforward manipulations we get the following 
asymptotic expansion
$$
\Omega_{0}(E)\sim\sin{\left(\pi d\over 2\right)}
e^{\beta_{H}E}\sum_{n=0}^{\infty} (-1)^{n+1}
a_{n}^{(1)} E^{-n-{d\over 2}-1}\int_{0}^{\eta E}{dy\over\pi}y^{n+a}e^{-y}
+{\cal O}(e^{cE})
$$
with $\eta=\beta_{H}-c$. Since $d=D-1$ is odd the prefactor is non-vanishing.

Up to here we have assumed that $c$ lies between the first and the second
branch point. However we can push $c$ further to the left and place it 
between $\beta_{M}$
and $\beta_{M+1}$. In this case we can split the contours $C_{\pm}$ into
the intervals $(\beta_{k+1},\beta_{k})$ ($k=1,\ldots,M$) and compute the
integral over each piece by expanding\foot{This can always be done since
the radius of convergence of the series expansion around $\beta_{k}$
will be equal to the distance to the closest singularity, 
 which in our case
is $\beta_{k+1}$ for every $k=1,2,\ldots$} 
${\cal Z}_{0}$ around the corresponding
$\beta_{k}$. Each 
expansion has exactly the same structure as \expaa, but now with
different coefficients $a^{(i)}_{n}$, $b^{(i)}_{n}$; again only the
multivalued parts contribute. Repeating the same steps as above on each
piece we finish with 
\eqn\aaw{
\Omega_{0}(E)\sim\sin{\left(\pi d\over 2\right)}\sum_{k=1}^{M}
e^{\beta_{k}E}\sum_{n=0}^{\infty} (-1)^{n+1}
a_{n}^{(k)} E^{-n-{d\over 2}-1}\int_{0}^{\eta_{k} E}{dy\over\pi}y^{n+a}e^{-y}
+{\cal O}(e^{cE})
,}
where $\eta_{k}=\beta_{k}-\beta_{k+1}$ ($k=1,\ldots,M-1$) and $\eta_{M}=
\beta_{M}-c$.

If we are only interested in the leading behaviour around each singular
point $\beta_{k}$, we conclude from the preceding analysis that to a leading
critical behaviour 
${\cal Z}(\beta)\sim (\beta-\beta_{k})^{\alpha}$ will correspond a
leading asymptotic term in the density of states:  
$$
\Omega(E)\sim {e^{\beta_{k}E}\over E^{\alpha+1}}+{\rm subleading\,\,terms}. 
$$
A term like this will contribute to the specific heat with
$$
{1\over c_{V}}\sim 
-{\alpha+1\over E}\left(\beta_{k}-{\alpha+1\over E}\right)^{2}
,$$
which will be negative unless $\alpha<-1$.

\appendix{C}{Higher-order Diagrams in Open String Theory}

Closed string theory is perturbatively defined in terms of 
a two-dimensional quantum field theory, defined on a closed,
orientable Riemann surface. Although the actual computation
of string processes beyond one loop is extremely complicated,
the mathematical simplicity of some general properties of
closed Riemann surfaces makes it possible to extract some
information about the behaviour of the contributions from 
Riemann surfaces of higher genus; for example one can write
explicitly the sum over classical embeddings in any
correlation function or the behaviour of
the string measure in some corners of the moduli space
at arbitrary genus.

All this simplicity seems to be lost in open string theory,
where also bordered Riemann surfaces contribute; the moduli
space becomes enlarged since  we can now deform both handles
and boundaries. There is, however, a way in which one can
relate open string diagrams to much simpler closed string
ones and thus make use of standard procedures to get
information about higher orders in open string perturbation 
theory. In this Appendix we will review the technique of
doubled bordered Riemann surfaces \refs\dob, \blau, 
\rgut,  and apply
it to prove that there are no corrections to the one-loop 
critical temperature in open bosonic
string perturbation theory as computed from the temperature-dependent
divergences associated with the boundary of the moduli space.

Given a genus-$h$ Riemann surface $\Sigma$ with $N$ boundaries, there
is a canonical way of getting a compact Riemann surface 
$\bar{\Sigma}$  of genus
$2h+p-1$,  by reflection of the original surface
with respect to its boundaries. Then
 $\Sigma$ can be obtained back from
its double $\bar{\Sigma}$ by dividing it by an antiholomorphic 
involution $I_{*}$ acting on the canonical homology basis of 
$\bar{\Sigma}$ as
$I_{*}A_{i}=\Gamma_{ij}A_{j}$ and $I_{*}B_{i}=-\Gamma_{ji}B_{j}
$,
where $\Gamma_{ij}\in {\bf Z}$ and $\Gamma^{2}=1$. The expression of $\Gamma_{
ij}$ can be simplified by the following choice of the canonical homology basis:
let $a_{\alpha}$, $b_{\alpha}$ ($\alpha=1,\ldots,h$) be the homology cycles 
of the handles already on $\Sigma$ and  $\tilde{a}_{\alpha}$, 
$\tilde{b}_{\alpha}$ their reflection with respect to the 
boundaries. In addition we also have $c_{l}$, $d_{l}$ ($l=1,\ldots,N-1$),
the cycles of $\bar{\Sigma}$ associated with the handles 
we are cutting 
to get $\Sigma$ back. In this basis the involution acts as $I_{*}c_{l}=c_{l}$,
$I_{*}d_{l}=-d_{l}$, $I_{*}a_{\alpha}=\tilde{a}_{\alpha}$ and 
$I_{*}a_{\alpha}=\tilde{a}_{\alpha}$. Therefore $\Gamma$ has the following
antidiagonal form
$$
\Gamma = \left(\matrix{ 0 & 0 & {\bf 1} \cr 0 & {\bf 1} & 0 \cr
{\bf 1} & 0 & 0}\right)
$$ 
The antiholomorphic action of $I_{*}$ on the doubled surface $\bar{\Sigma}$
means that its period matrix $\tau$ has to satisfy 
$\tau = -\Gamma^{\rm t}\bar{\tau}\Gamma $;
with our choice for $\Gamma$ this means
\eqn\period{
\tau = \left(\matrix{a & b & c \cr b^{\rm t} & it & -\bar{b}^{\rm t} \cr
-\bar{c} & -\bar{b} & -\bar{a} }\right)
}
where $a^{\rm t}=a$, $t$ is real and $c$ is anti-hermitian
($c=-{\bar{c}^{\rm t}}$).

If we want to compute scattering amplitudes for open strings from 
the doubled surface, we need
the open string measure in terms of the string measure on 
$\bar{\Sigma}$.  
The partition function for the closed bosonic string at genus $2h+N-2$ can be
written as \refs\dokph\
\eqn\closed{
\Lambda_{2h+N-2
}=\int_{{\cal M}(\bar{\Sigma})} d({\rm WP})_{\bar{\Sigma}} 
(\det (P^{+}P)_{
\bar{\Sigma}})^{1/2}
\left[{8\pi^{2}\over \int_{\bar{\Sigma}} d^{2}\xi \sqrt{g}}
\det{}'\Delta_{\bar{\Sigma}}\right]^{-13}\sum_{\rm solitons} e^{-S_{c\ell}
^{\rm closed}}
,}
where $d({\rm WP})$ is the Weil--Peterson measure defined in terms of the
quadratic and Beltrami differentials,  
and the sum over classic configurations can be written as
$$
\sum_{\rm solitons} e^{-S_{c\ell}^{\rm closed}(\vec{N},\vec{M})} =
\sum_{\vec{N},\vec{M}} e^{-{\beta^{2}\over 4\pi\alpha^{'}}
(\bar{\tau}\vec{N}-\vec{M})^{\rm t}({\rm Im\,}\tau)^{-1}
(\tau\vec{N}-\vec{M})}
.$$
On the other hand the open string path integral over $\Sigma$ 
is formally
\eqn\open{
\Lambda^{\rm open}_{h,N}=\int_{{\cal M}(\Sigma)} d({\rm WP})_{\Sigma}
(\det (P^{+}P)_{
\Sigma})^{1/2}
\left[{8\pi^{2}\over \int_{\Sigma} d^{2}\xi \sqrt{g}}
\det{}'\Delta_{\Sigma}\right]^{-13}\sum_{\rm solitons} e^{-S_{c\ell}^{\rm open}}
.}
The mapping between the determinants in \open\ and \closed\ has been 
computed in \refs\blau\ with the result
\eqn\facc{
\eqalign{\det (P^{+}P)_{\Sigma} &= (\det(P^{+}P)_{\bar{\Sigma}})^{1/2} \cr
{\det{}'\Delta_{\Sigma} \over \int_{\Sigma} d^{2}\xi \sqrt{g}} &=
\left({\det{}'\Delta_{\bar{\Sigma}} \over \int_{\bar{\Sigma}} d^{2}\xi 
\sqrt{g}}\right)^{1/2}R_{\bar{\Sigma},I}^{\mp 1/2},} 
}
where  $\mp$ is for Neumann/Dirichlet boundary conditions and
$R_{\Sigma,I}$ is defined by
$$
R_{\bar{\Sigma},I}=\det\left[(1+\Gamma){\rm Im\,}\tau+(1-\Gamma)
({\rm Im\,}\tau)^{-1}\right].    
$$
The integration region ${\cal M}(\Sigma)$ is a real submanifold of the 
moduli space ${\cal M}(\bar{\Sigma})$ and therefore the square roots 
are {\it holomorphic} square roots.

Special care is needed in the case of the Weil--Peterson measure. Using
the complex structure $J$ on $\bar{\Sigma}$, which anticommutes with $I_{*}$,
one can write a base of quadratic differentials \refs\blau\  $\{S_{i},J S_{i}
\}$, where $I_{*}S_{i}=S_{i}$. In the same way this can be done also for a
base of Beltrami differentials. Following the details in \refs\blau\ one
finds that
$$
d\,({\rm WP})_{\bar{\Sigma}}= d\,({\rm WP})_{\Sigma}\wedge J d\,
({\rm WP})_{\Sigma} , 
$$
where the measure for the open string is explicitly written as
$$
d\,(WP)_{\Sigma}={\det \langle S_{i}|\mu_{j}\rangle \over 
(\det \langle S_{i}|S_{j}\rangle)^{1/2}}\prod_{l=1}^{3g-3} dm_{l}
$$
with $\mu_{i}$ the Beltrami differentials.  

Finally we have to take care of the solitonic sum in \open. Since $I_{*}$ 
acts on the homology cycles of $\bar{\Sigma}$, the same will be true for
the winding numbers of these cycles around the compact target dimension
$$
I_{*}N_{i}=\Gamma_{ij}N_{j}, \hskip 1cm I_{*}M_{i}=-\Gamma_{ji}M_{j}. 
$$
Because we are dividing by $I_{*}$, we are interested in retaining only
those winding configurations that are symmetric under the involution.
This implies that we have to truncate the solitonic sum down to those
winding numbers satisfying $\Gamma \vec{N}=\vec{N}$ and $\Gamma^{\rm t}
\vec{M}=-\vec{M} $:  
\eqn\sol{
\sum_{\rm solitons} e^{-S_{cl}^{\rm open}}=
\sum_{\vec{N},\vec{M}} \delta(\Gamma \vec{N}-\vec{N})
\delta(\Gamma^{\rm t}\vec{M}+\vec{M})
e^{-{1\over 2}S_{c\ell}^{\rm closed}(\vec{N},\vec{M})} 
.}
The factor $1\over 2$ in front of the classical action arises because the
area of the doubled surface is twice the area of the original bordered
surface $\Sigma$ and therefore the classical action for open strings is
half that for the auxiliary   closed string theory.

By using the
base in which $\Gamma$ is antidiagonal,  
it is straightforward to check that the winding numbers of the $A_{i}$
and $B_{i}$ cycles
are respectively of the form
$$
\eqalign{
\vec{N}&=(n_{1},\ldots,n_{h};q_{1},\ldots,q_{p-1};n_{1}\ldots,n_{h}) \cr
\vec{M}&=(m_{1},\ldots,m_{h};0,\ldots,0;-m_{1},\ldots,-m_{h}). 
}$$
An overall change of sign in $\Gamma$ amounts to an interchange of $A$ with
$B$ cycles (modular transformation) and therefore the structure of the
winding numbers also gets interchanged.

In the remainder of this Appendix we will use the techniques 
reviewed so far to compute the temperature-dependent divergences
associated with the boundary of the moduli space of higher-order
Riemann surfaces with boundaries.
Let us assume that we have a genus-$h$ Riemann
surface with $N$ boundaries attached to a $p$-dimensional 
D-brane\foot{In what follows we will consider that
the compactified Euclidean time is always a Neumann direction
and therefore $-1<p\leq 25$. Later on we will describe how
to construct the D-instanton case $p=-1$ from our expressions.}. 
Using what we have learned, we can write the formal
contribution of such a diagram in terms of determinants 
evaluated on the doubled surface as
$$
\Lambda_{h,N}=\int_{{\cal M}(\Sigma)} d({\rm WP})_{\Sigma}
\sqrt{(\det (P^{+}P)_{\bar{\Sigma}})^{1/2}
\left[{8\pi^{2}\det{}
^{'}\Delta_{\bar{\Sigma}}\over \int_{\bar{\Sigma}}d^{2}\xi
\sqrt{g}}\right]^{-13}}R_{\bar{\Sigma},I}^{p-12\over 2}
\sum_{\rm solitons} e^{-S_{c\ell}}
,$$
where the square root is understood as a holomorphic 
square root.
Divergences in $\Lambda_{h,N}$ will be associated with 
the boundary of the moduli space ${\cal M}(\Sigma)$; since
we are interested in temperature-dependent singularities
we can obviate any singular behaviour triggered by 
zero-momentum states. This means that we can focus our attention
in two of the components of the boudary of ${\cal M}(\Sigma)$:
the shrinking of a boundary or the degeneration of a closed 
string handle.

Let us go first to the degeneration of a boundary. From 
\period\ we see that this corresponds, for example,
to the limit ${\rm Im\,}\tau_{h+1,h+1}=t_{11}
\rightarrow \infty$. It can  easily be checked that
none of the entries of $({\rm Im\,}\tau)^{-1}$ diverges in
that limit, so from \sol\ we conclude that
\eqn\sst{
\sum_{\rm solitons} e^{-S_{c\ell}}\sim
1+2\,e^{-{\beta^{2}t_{11}\over 8\pi\alpha^{'}}}
+\ldots
}
Having analysed the sum over classical vacua we turn
to the path integral measure. Now we take advantage of having
written the open string measure as, essentially, the square root
of the closed string measure on the doubled surface and,  
 using Belavin--Knizhnik theorem \rbn, we extract the limit  
$$
\sqrt{
(\det(P^{+}P)_{\bar{\Sigma}})^{1/2}\left[{8\pi^{2}\det{}
^{'}\Delta_{\bar{\Sigma}}\over \int_{\bar{\Sigma}}
d^{2}\xi \sqrt{g}}\right]^{-13}} \sim
|e^{2\pi i\tau_{h+1,h+1}}|=e^{2\pi t_{11}}
.$$
This limit, together with the large $t_{11}$ expansion of
the solitonic part, implies that the critical temperature is
$\beta_{c}=4\pi\sqrt{\alpha^{'}}$, in perfect accordance with
the one-loop result.

To complete the computation we have to study those divergences 
associated with the degeneration of a closed string handle.
Now, however, because of the form of the period matrix \period,  
this limit corresponds to the degeneration
of two handles in $\bar{\Sigma}$, those associated with the
entries $\tau_{11}$ and $\tau_{h+N,h+N}$. Again it is 
very easy to find the asymptotic behaviour of the solitonic part
in the limit ${\rm Im\,}\tau_{11}={\rm Im\,}\tau_{h+N,h+N}\equiv a
\rightarrow \infty$ as
$$
\sum_{\rm solitons} e^{-S_{c\ell}}\sim 1+2\,e^{-{\beta^{2}a\over
4\pi\alpha^{'}}}+\ldots
$$
Notice the  additional factor of 2 in the exponent with respect to 
\sst,  due to the fact that we are degenerating two handles at a time.
The Belavin--Knizhnik
 theorem comes again to the rescue in the computation
of the limit of the string measure to find
$$
\sqrt{
(\det(P^{+}P)_{\bar{\Sigma}})^{1/2}\left[{8\pi^{2}\det{}^{'}
\Delta_{\bar{\Sigma}}\over \int_{\bar{\Sigma}}d^{2}\xi\sqrt{g}}
\right]^{-13}} \sim e^{4\pi a}
,$$
where again we have to keep in mind that we have two soft handles.
Thus comparing the two exponential factors we come to the same value
of the critical inverse temperature, $\beta_{c}=4\pi\sqrt{\alpha^{'}}$.

We close our discussion with some remarks on the D-instanton
case. As we already pointed out, in our framework the compact direction
always has Neumann boundary conditions and this excludes the degenerate
case $p=-1$. To recover the D-instanton we must begin with the D-particle
($p=0$) and perform a Poisson resummation of the sum over classical vacua
followed by a T-duality $\beta\rightarrow 4\pi^{2}\alpha^{'}/\beta$ 
(for a discussion including higher-genus surfaces, see \refs\renbor). 
Following the same line of reasoning, we find the critical temperature
$\beta_{c}=2\pi\sqrt{\alpha^{'}}$ for all diagrams in perturbation theory.

We have concluded that higher-order diagrams do not modify the one-loop
value of 
the critical length $\beta_{c}$ at which tachyon-triggered singularities
appear; in this sense, the situation is pretty  similar to that for
closed
 strings \refs\ralor, \raloros.
 Whether or not this implies a non-renormalization
of the
 Hagedorn temperature is much more difficult to decide, since non-trivial
target-space effects may take place
 near the Hagedorn transition. In any case
the situation is far from settled.

\listrefs
\bye